\documentclass[aps,prc,superscriptaddress,twocolumn,showpacs]{revtex4}
\usepackage{graphicx}
\usepackage{dcolumn}

\usepackage{color} 

\usepackage{amsfonts}
\usepackage{amssymb}
\usepackage{amsmath}
\usepackage{amsxtra}

\begin{document}

\title{Isobaric-multiplet mass equation in a macroscopic-microscopic approach}

\author{O.~Klochko}
\email[]{zcemokl@ucl.ac.uk}
\affiliation{The Haberdashers’ Aske’s Boys’ School, Butterfly Lane, 
Elstree, Hertfordshire, WD6 3AF, UK}
\affiliation{Department of Mechanical Engineering, University College London, Torrington Place, London WC1E 7JE, UK}

\author{N.~A.~Smirnova} 
\email[]{smirnova@cenbg.in2p3.fr}
\affiliation{CENBG (CNRS/IN2P3 - Universit\'e de Bordeaux), 33175 Gradignan cedex, France }

\date{\today}

\bibliographystyle{prsty}

\begin{abstract}
We study the $a$, $b$ and $c$ coefficients of the isobaric-multiplet mass equation using a macroscopic-microscopic approach 
developed by P. M\"{o}ller and collaborators 
[At. Data Nucl. Data Tables {\bf 59},  185  (1995); {\it ibid} {\bf 109-110},  1  (2016)].
We show that already the macroscopic part of the finite-range liquid-drop model (FRLDM) describes 
the general trend of the $a$ and $b$ coefficients relatively well, 
while the staggering behavior of $b$ coefficients for doublets and quartets can be understood in terms of 
the difference of average proton and neutron pairing energies.
The sets of isobaric masses, predicted by the full macroscopic-microscopic approaches, are used to 
explore the general trends of IMME coefficients up to $A=100$. 
We conclude that while the agreement for $a$ coefficients is quite satisfactory, the full approaches have less sensitivity 
to predict the IMME $b$ and $c$ coefficients in detail.
The best set of theoretical $b$ coefficients, as given by the modified macroscopic part of the FRLDM, 
is used to predict masses of proton-rich nuclei based on the known experimental masses of neutron-rich mirror partners,
and subsequently to investigate their one- and two-proton separation energies in proton-rich nuclei up to the $A=100$ region. 
The estimated position of the proton-drip line is in fair agreement with known experimental data.
These masses are important for simulations of the astrophysical $rp$-process.

\end{abstract}


\maketitle

\section{Introduction}

The concept of isospin was introduced by Heisenberg in 1939~\cite{Heisenberg1932}. 
Since then, it represents a very useful paradigm in nuclear and particle physics, providing more beauty and simplification 
in theoretical modeling and interpretation of hadron or nuclear properties.

According to the isospin formalism,  a nucleon is an isospin $t=1/2$ baryon, 
with a neutron and a proton being assigned $t_z = 1/2$ and $t_z = -1/2$, respectively.  
The three Cartesian components of the isospin operator $\hat{\mathbf{t}}$ obey the $su(2)$ commutation relations,
\begin{equation}
\label{commutator}
[\hat t_i,\hat t_j] = {\rm i} \varepsilon_{ijk}\hat t_k \,
\end{equation}
and $ [\hat{\mathbf{t}}^2,\hat t_i] =0$, where $i,j,k$  run over $x,y,z$, and 
$\hat{\mathbf{t}}^2 =  \hat t^2_x + \hat t^2_y + \hat t^2_z $. 

\begin{figure}
 \centering
  \includegraphics[width=0.4\textwidth]{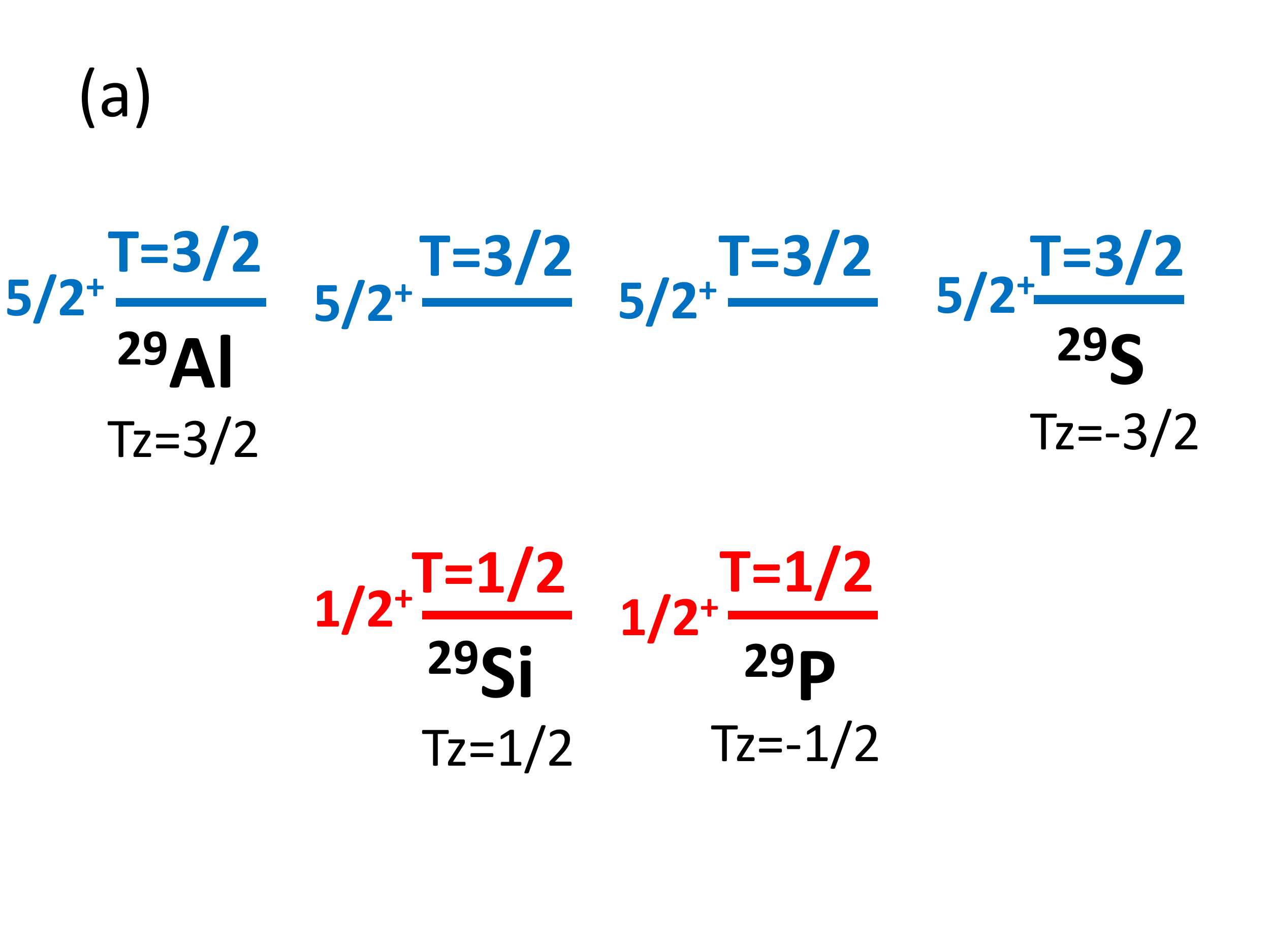}\\[2mm]
  \includegraphics[width=0.4\textwidth]{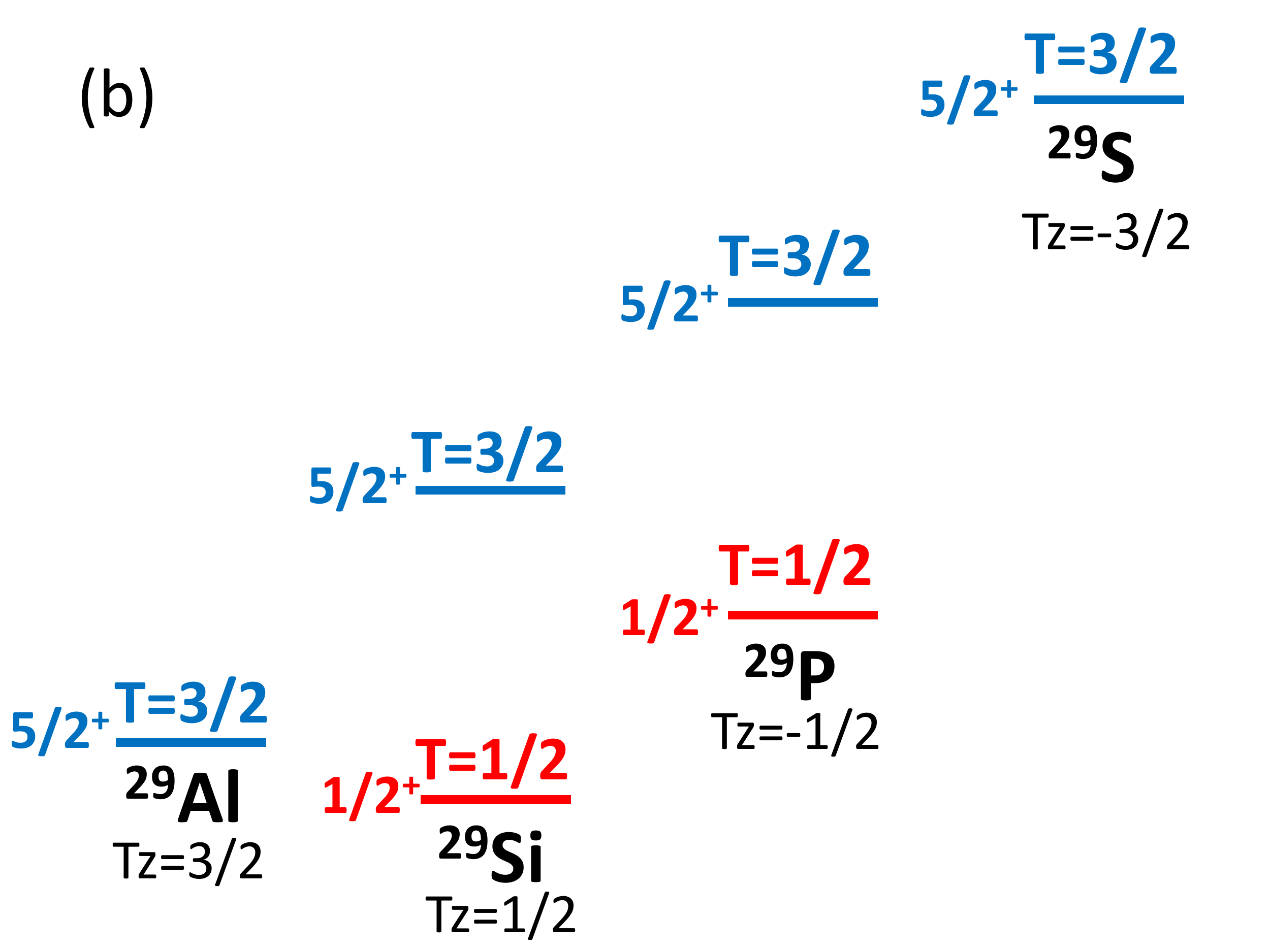} 
  \caption{\label{fig:A=29} (Color online)  The lowest $T=1/2$ and $T=3/2$ isobaric multiplets in $A=29$:
(a) schematic illustration of degenerate multiplets in the absence of charge-dependent interactions;
(b) realistic situation obtained using experimental masses excesses of the corresponding nuclei. 
See text for details.} 
\end{figure}

In the absence of electromagnetic interactions and under the assumption of equal proton and neutron masses, 
a Hamiltonian of a nucleus would commute with the many-body
isospin operator, $\hat{\mathbf{T}}=\sum_{n=1}^A \hat{\mathbf{t}}(n)$, and its eigenstates would represent degenerate 
isospin multiplets $|T T_z \rangle $, characterized by two quantum numbers, $T$ and $T_z$, with $T_z = -T, -T+1, \ldots, T$
(see Fig.~\ref{fig:A=29}(a) for illustration). 
The members of the isobaric multiplets are called {\em isobaric analogue states}. 
For a given nucleus $T_z=(N-Z)/2$ and the isospin quantum number can take values
for different states of $T=|T_z|, |T_z|+1, \ldots , A/2$. 

The presence of the Coulomb interaction between protons and possible small isospin-nonconserving forces of nuclear origin 
break the isospin symmetry.
Assuming a two-body nature of charge-dependent forces and conservation of the isospin of a two-nucleon system, 
Wigner noticed~\cite{Wigner58} that those isospin-symmetry breaking operators will be a combination of an isoscalar, 
an isovector and an isotensor operators.
Estimating the splitting of the isobaric multiplets in lowest order perturbation theory due to the expectation value of such an operator, Wigner showed~\cite{Wigner58} that isobaric multiplets will be split according to a quadratic equation in $T_z$, called  {\it isobaric multiplet mass equation} (IMME),
\begin{equation}
\label{eq:IMME}
M(\eta ,T,T_z) = a(\eta ,T) + b(\eta ,T) T_z + c(\eta ,T) T_z^2 \, .
\end{equation}
Here, $M(\eta ,T,T_z)$ refers to the atomic mass excess,
$\eta = (A,J^{\pi }, N_{exc},\ldots )$ denotes all other than $T$ quantum numbers (atomic mass number, spin and parity,
number of an excited state, $\ldots $), which are
required to label a quantum state of an isobaric multiplet, whereas $a$, $b$ and $c$ are coefficients.
As an example, the splittings of the lowest $T=1/2$ and $T=3/2$ isobaric multiplets in $A=29$ are shown in Fig.~\ref{fig:A=29}(b).

There has always been a lot of interest in the IMME, which proved to work for a great number of quartets and quintets as well, serving thus as a ground for nuclear mass models~\cite{JaneckeMasson}.
The properties of the coefficients -- their global trends and a specific staggering phenomenon --
have been studied  theoretically since 1960s~\cite{Janecke1966a,Hecht1966,Hecht1967,Janecke1969} up to the present  
(see, e.g., Refs.~\cite{LamPRC2013,Kaneko2013,ADNDT2013,MacCormick2014,Tu2014,Ormand2017,Baczyk2017,Fu2018} for the very recent work).
A particular attention has been paid to experimental search (e.g., ~\cite{Brodeur2017} and references therein) and 
theoretical interpretation of cubic or quartic terms in the IMME due to isospin mixing or to manifestations 
of the charge-dependent many-body forces~\cite{Henley_Lacy69,Janecke69,BertschKahana70,Benenson1979,SiBr11,Dong2018,Dong2019}.

The advent of modern radioactive ion beam facilities, progress in mass measurements and particle detection techniques allowed to access multiplets with more and more short-lived members. 
The most recent compilations~\cite{ADNDT2013,MacCormick2014} of IMME coefficients 
contain an impressive amount of highly precise data which provides general trends of the IMME coefficients and sometimes staggering effects as a function of $A$, highlighting specific properties of the nucleon-nucleon interaction and giving certain hints on 
existing shell effects.

%

Recent achievements in microscopic many-body theory allow to calculate IMME parameters using both phenomenological and microscopic effective nucleon-nucleon forces~\cite{OrBr89,Ormand96,Ormand97,Kaneko2013,LamPRC2013,Ormand2017, Baczyk2017,Baczyk2018,Dong2018,Dong2019}. 
In empirical approaches, experimental databases are often used to establish the strength in isovector and isotensor channels of an effective nuclear force.
However, to provide a uniform description of the whole region of data from $A=5$ to $A\sim 100$ nuclei is still challenging.

A particular high interest to masses of proton-rich nuclei up to $A\approx 100$ is also due to their significance 
for simulations of the astrophysical $rp$-process which powers type-I X-ray bursts~\cite{Schatz1998}.
These are periodic events which occur in binary systems consisting of a rapidly rotating neutron star, 
accreting hydrogen- and helium-rich material from its companion, typically, a main-sequence star. 
At high temperature and pressure conditions, the burst is ignited, powered by the explosive burning
of hydrogen and helium~\cite{WallaceWoosley1981}. 
The radiative proton capture reactions are the most important reaction sequence, which in competition with the $\beta^+$ decay
is responsible for synthesis of proton-rich nuclides up to the $A=100$ region.
Nuclear physics input on masses, one- and two-proton separation energies, and $\beta $-delayed processes are thus of great importance
for X-ray burst models for reliable calculations of light curves and predictions of final abundances 
(see e.g. Ref.~\cite{Cyburt2016} for the current status of state-of-the-art modelizations).
This leads to enormous experimental and theoretical efforts to provide astrophysicists with unknown masses of proton-rich nuclei~\cite{GraweRPP2007}.
Because of the complexity of the nuclear many-body problem and extreme conditions,
various approaches achieve a consensus on some proton drip-line nuclei, while disagree on others.
Thus, there is still need for more understanding and more accurate theoretical predictions.

In this work, we propose to construct theoretical $a$, $b$ and $c$ coefficients and study their properties
using a global macroscopic-microscopic approach developed 
during a few decades by P. M\"{o}ller and collaborators~\cite{MoellerNixADNDT1995,MoellerADNDT2016}.
This framework, designed for global nuclear mass calculations, consists of two parts: a macroscopic part and a microscopic one.
Two different models have been used throughout the years to represent a macroscopic contribution, 
giving the names to the corresponding approaches:
(i) the finite-range liquid-drop model (within the FRLDM), and (ii) the finite-range droplet model (within the FRDM). 
The former is a far more refined version of the uniformly-charged liquid drop model,
while the latter includes in addition a few high-order terms in $A$.
In both cases, a similar microscopic term has been used, consisting of a shell-plus-pairing correction. 
The parameters of the models have been thoroughly adjusted to a large variety of known data on nuclear masses.
The details of the approach and parametrization strategy can be found in the original work~\cite{MoellerNixADNDT1995,MoellerADNDT2016}.

Although phenomenological, the macro-microscopic models have robust grounds and are optimized 
to describe a few thousands of nuclear masses (around 2150) 
with very small root-mean-square (rms) error of around 0.56~MeV (0.66~MeV) for FRDM (FRLDM)~\cite{MoellerADNDT2016}.
The models have been applied mainly to predict masses of (super)heavy elements and fission barriers.
In this work, we apply this approach rather to nuclei along the $N=Z$ line and study the general trends and 
fine structure of the deduced IMME coefficients as a function of the mass number.
Some preliminary expressions for $b$ and $c$ coefficients obtained from the macroscopic part of the FRLDM 
have already been deduced in Refs.~\cite{Ormand97,LamPRC2013} and applied for either $sd$ or $pf$ shell nuclei. 
In the present study, we go beyond and we obtain all IMME coefficients from the macroscopic part of FRLDM and 
from the full macroscopic-microscopic approaches, and we apply it to the whole range of experimental data
(excluding only very light nuclei).
Starting with a well-known uniformly charged liquid-drop model in Section III, 
we introduce the macroscopic part of the FRLDM from Ref.~\cite{MoellerADNDT2016} 
as a more involved and improved way of global description of the IMME coefficients (Section IV).
We discuss general trends provided by the model, as well as we show that the staggering effect of $b$ coefficients can be described by different proton and neutron pairing energy parameters.
In Section V, we explore the predictions given by the full FRLDM and FRDM~\cite{MoellerADNDT2016}.
Finally, in Section VI, we use the macroscopic part of the FRLDM, which provides the ``best'' set of $b$ coefficients,  
to calculate masses of proton-rich nuclei in the vicinity of the proton drip-line from Mn isotopes up to Te isotopes.
We discuss the estimated one- and two-proton separation energies in the context of the available experimental data.
The last section summarizes the conclusions and outlines the perspectives of this study.


\section{Experimental determination of IMME coefficients}

Starting from Eq.~(\ref{eq:IMME}), one can express  $a$, $b$ and $c$ coefficients for a given isobaric multiplet  $(\eta, T)$ 
in terms of mass excesses of its members. Denoting $M(\eta, T, T_z)\equiv M_{T_z}$, we get for doublets ($J^\pi, T=1/2$) that 
\begin{equation}
\label{ab-doublets}
\begin{array}{l}
a = (M_{1/2} +M_{-1/2})/2  \, ,\\
b =  M_{1/2} -M_{-1/2} \,, 
\end{array}
\end{equation}
while $a$, $b$ and $c$ coefficients for triplets ($T=1$) can be expressed as
\begin{equation}
\label{ab-triplets}
\begin{array}{l}
a =  M_0  \, ,\\
b =  (M_1 -M_{-1})/2 \,\\
c =  (M_1 +M_{-1})/2 - M_0 \,.
\end{array}
\end{equation}
It is obvious from these expressions, that ground state mass excesses only can serve to obtain $a$ and $b$ coefficients for ground state doublets and $b$ coefficients for the lowest triplets.
Generally, $M_0$ involves an excited state, which is situated at a particularly high excitation energy in an $N=Z$ even-even nucleus. 
This prohibits derivation of $a$ and $c$ coefficients for triplets from masses only for detailed comparison with the experimental data.

For quartets ($T=3/2$), quintets ($T=2$) or other high $T$ multiplets, where masses of more than three members are involved,
one determines $a$, $b$ and $c$ coefficients by a least-squares fit. 
For example, for quartets, the mass excesses of the isobaric analogues states are related by the IMME coefficients via the following system:
\begin{eqnarray}
M_{3/2} & = & a+\frac{3}{2} b + \frac{9}{4} c  \, ,\\
M_{1/2} & = & a+\frac{1}{2} b + \frac{1}{4} c  \, ,\\
M_{-1/2}& = & a-\frac{1}{2} b + \frac{1}{4} c  \, ,\\
M_{-3/2}& = & a-\frac{3}{2} b + \frac{9}{4} c  \, . 
\end{eqnarray}
The members with $T_z=\pm 1/2$ require knowledge of the excitation energy of those states.

In the present study, we approximate $b$ coefficients for $T>1$ multiplets by their relation to the Coulomb displacement energies between members with $T_z=T$ and $T_z=-T$~\cite{BentleyLenzi,Kaneko2013}:
\begin{equation}
\label{b-quartets}
b =  (M_{T} -M_{-T})/2T.
\end{equation}
This turns out to be sufficient for a discussion of the general trends and even of staggering phenomena.


\section{IMME coefficients from a uniformly charged liquid drop}

\begin{figure*}
 \centering
  \includegraphics[width=0.22\textwidth]{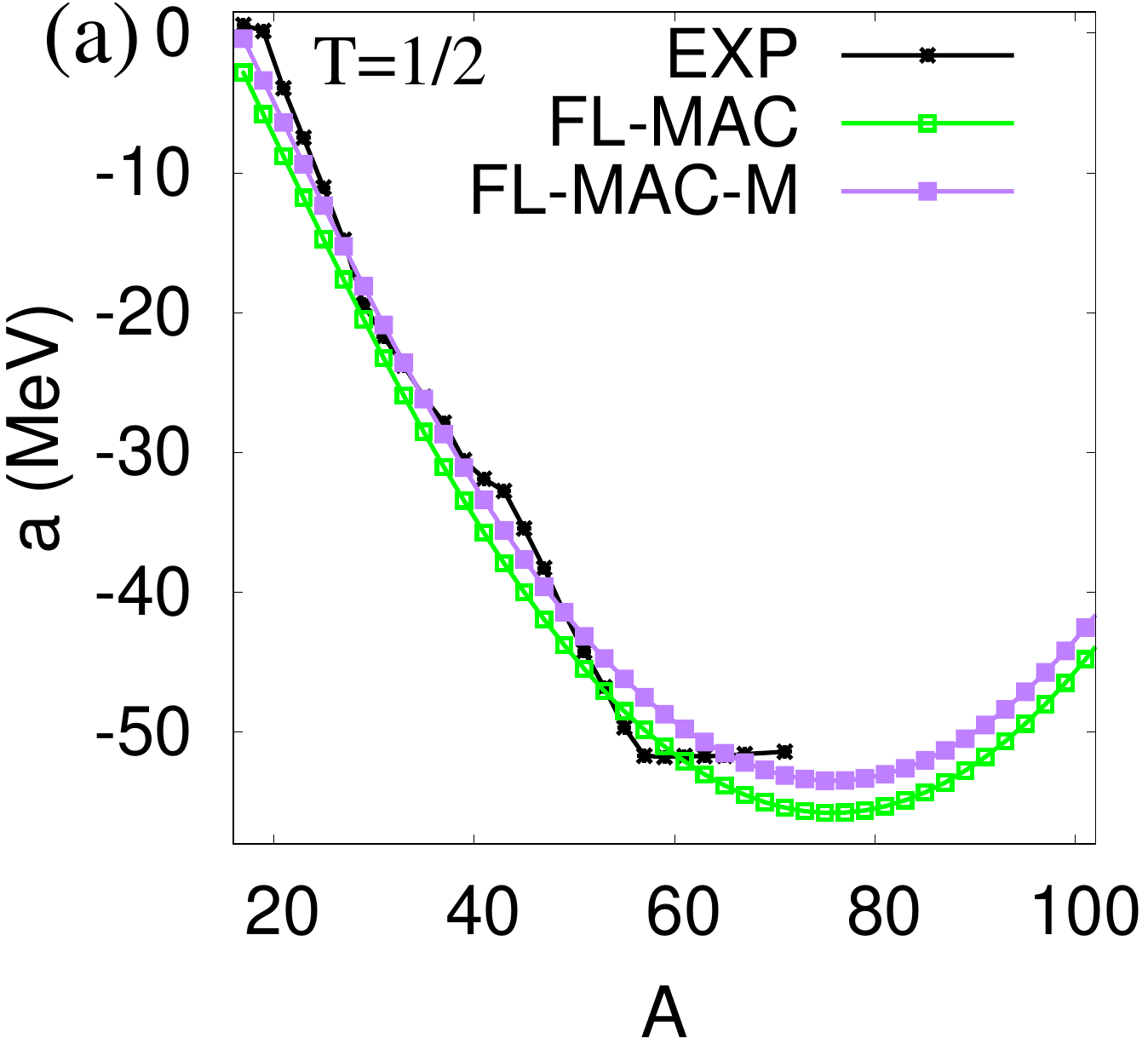} \hspace{1mm}
  \includegraphics[width=0.22\textwidth]{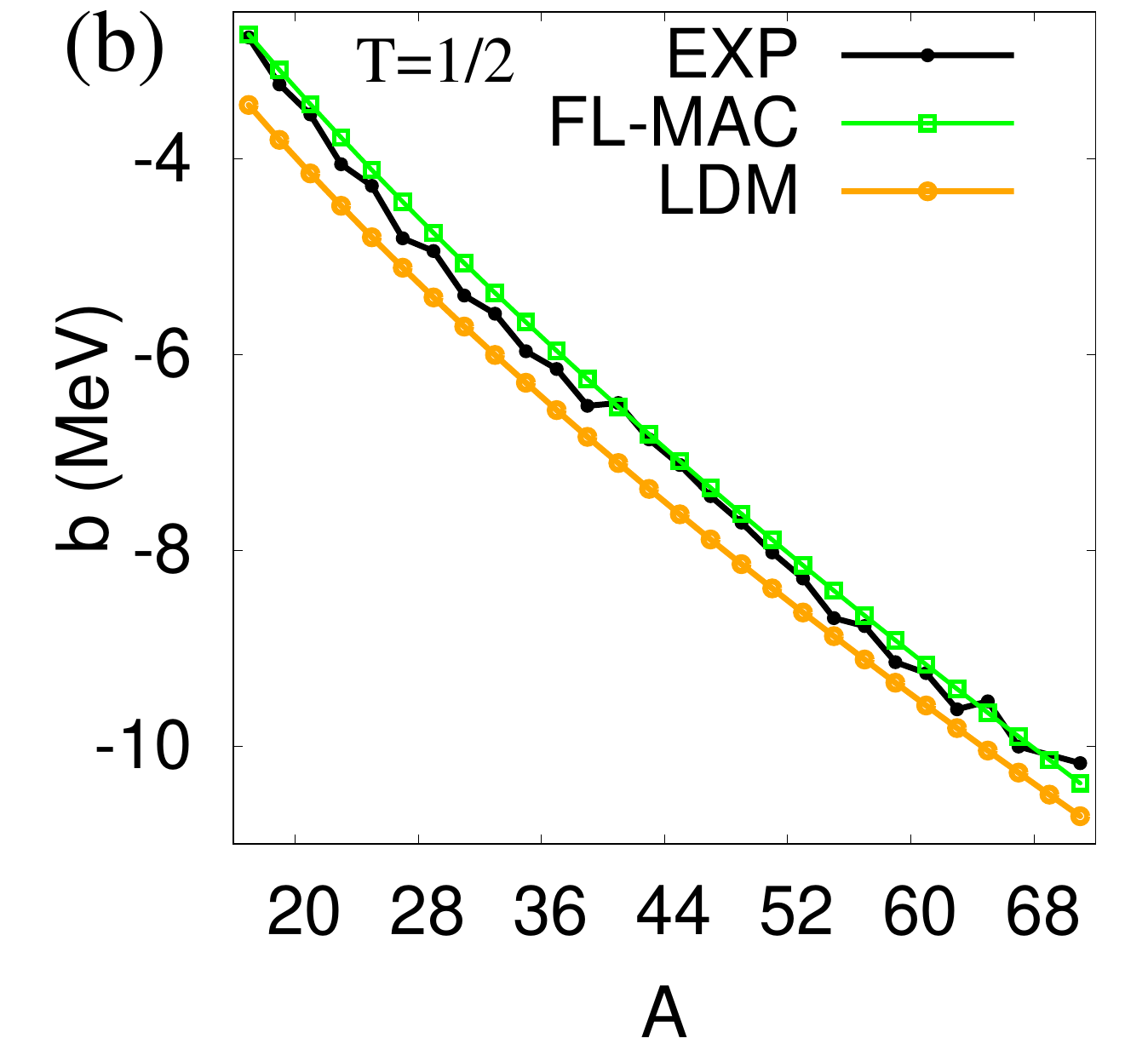} \hspace{1mm}
  \includegraphics[width=0.22\textwidth]{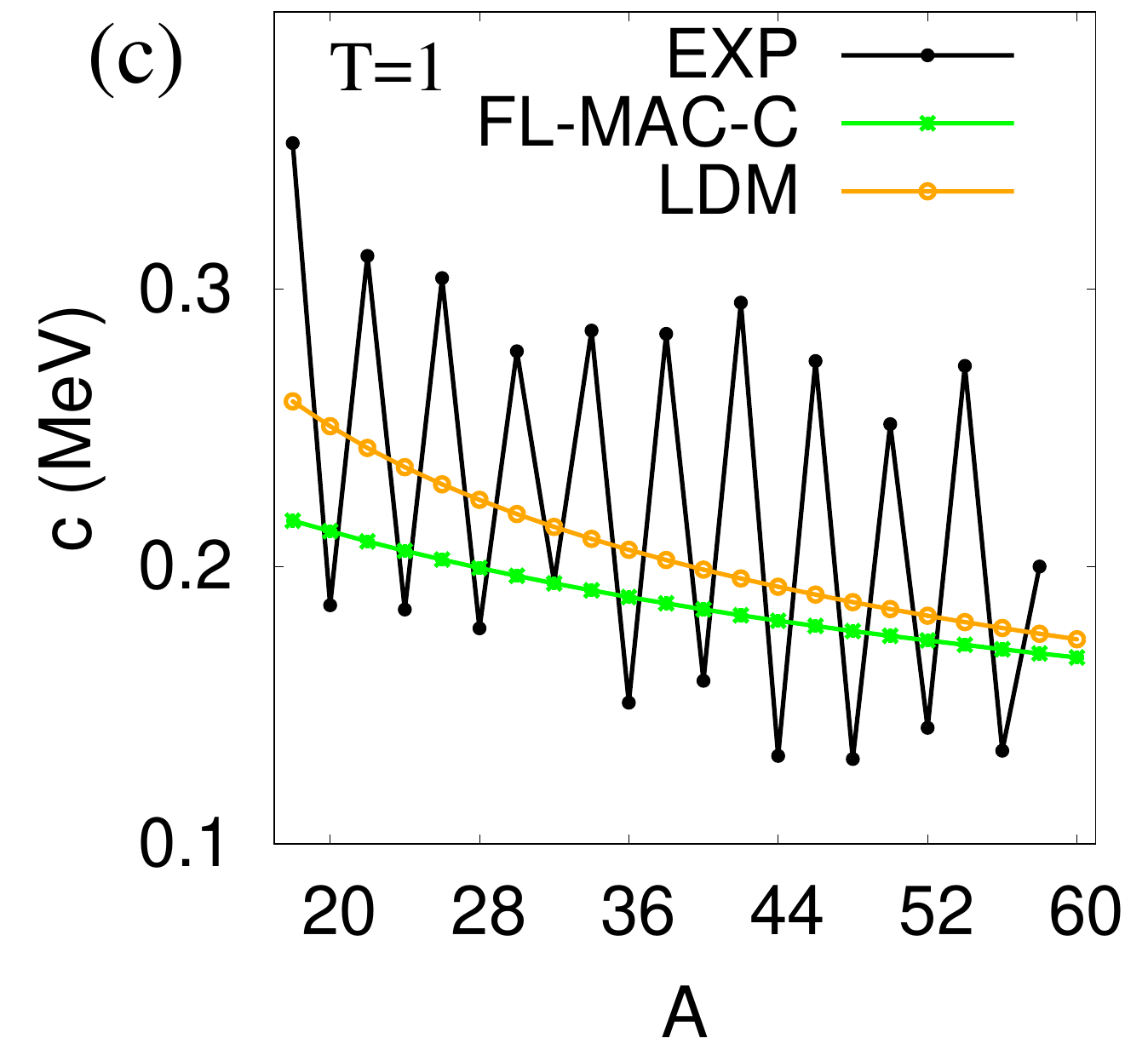} \hspace{1mm}
  \includegraphics[width=0.22\textwidth]{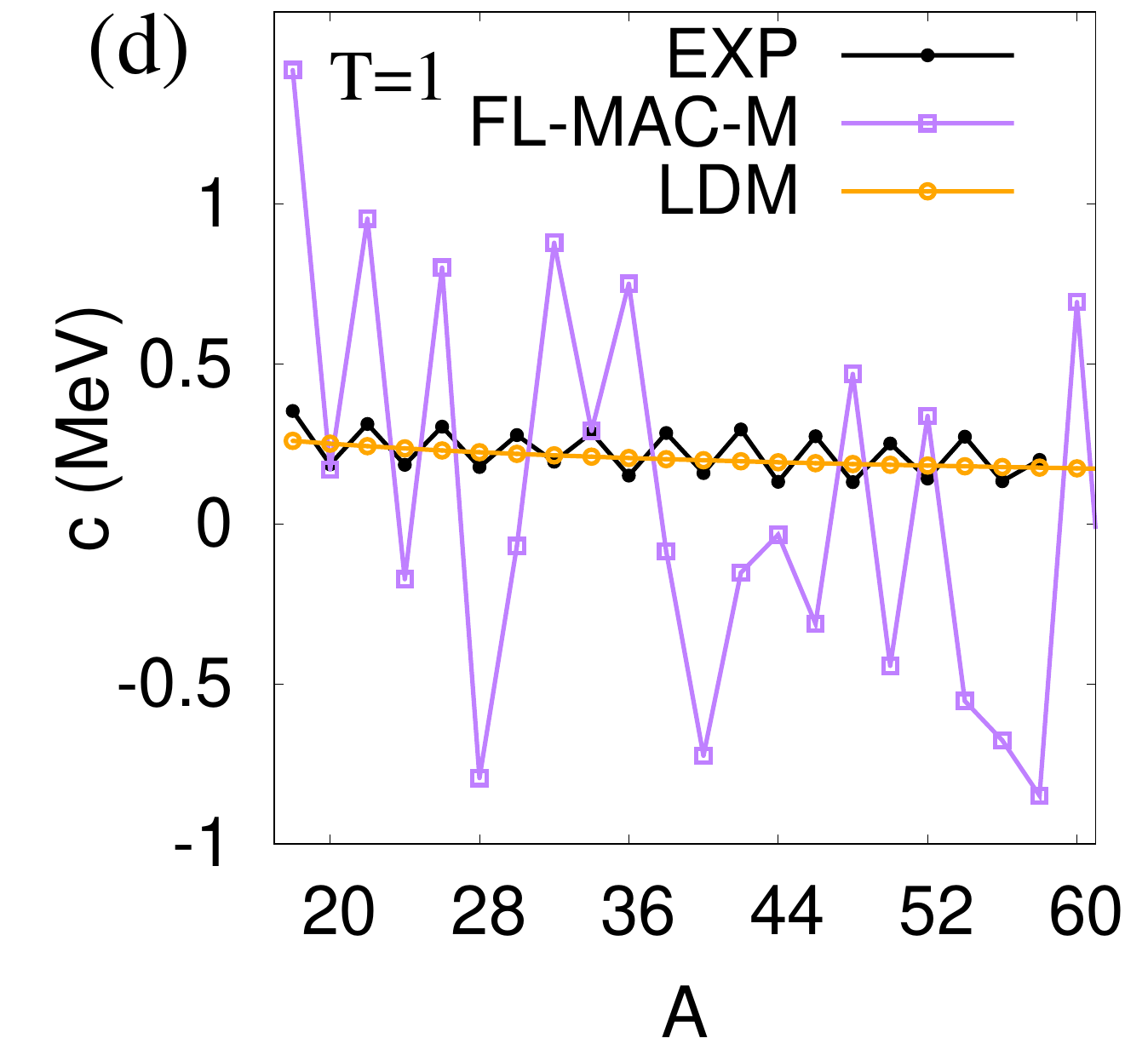}
  \caption{\label{fig:abc-trend} (Color online) Theoretical IMME $a$ and $b$ coefficients for doublets (panels (a) and (b), respectively) and $c$ coefficients for triplets (panels (c) and (d)) obtained from 
the macroscopic part of the FRDLM (FL-MAC) in comparison with experimental data (EXP) and liquid-drop model predictions (LDM). 
The calculations labeled as FL-MAC-M in panels (a) and (d) refer to the modified macroscopic part of the FRLDM, while calculations
labeled as FL-MAC-C on panel (c) designate a Coulomb approximation for $c$ coefficients.
Results from the mass models in the last panel have been supplemented by the experimental excitation energy in $T_z=0$ members. 
See text for further details.} 
\end{figure*}

We start with estimates of the contributions to $a$, $b$ and $c$ coefficients from the uniformly-charged sphere model~\cite{BetheBacher1936}. The total Coulomb energy of a uniformly charged spherical nucleus of radius
$R=r_0 A^{1/3}$ reads 
\begin{eqnarray}
\label{eq:chargedSphere}
	E_{coul} &= &\frac{3e^2}{5 R} Z(Z-1) \nonumber\\
	         &= &\frac{3e^2}{5 r_0 A^{\frac{1}{3}}} \left[\frac{A}{4}(A-2)+(1-A)T_z + T_z^2 \right],
\end{eqnarray}
giving rise to the following contributions to the IMME $a$, $b$ and $c$ coefficients~\cite{Benenson1979,BentleyLenzi}:
\begin{align}
\label{eq:chargedSphere_abc}
		a &= \frac{3e^2}{20 r_0} \frac{A(A-2)}{A^{1/3}} \, ,\\ 
		b &= -\frac{3e^2}{5 r_0 } \frac{(A-1)}{A^{1/3}} + \Delta_{nH}\, , \\ 
		c &= \frac{3e^2}{5 r_0 } \frac{1}{A^{1/3}} \, , 
\end{align}
where $e^2 = 1.44$~MeV$\cdot$fm and we use here the value of $r_0=1.27$~fm.
The quantity $\Delta_{nH}=782.346$~keV is the difference between the neutron ($M_n$) and hydrogen ($M_H$) mass excesses.
The smooth trends, given by these equations for $b$ and $c$ coefficients, are shown in Fig.~\ref{fig:abc-trend} (b,c), respectively 
(labeled as LDM -- liquid-drop model) in comparison with the experimental data on $b$ coefficients for doublets with $A=17-71$ and 
$c$ coefficients for triplets with $A=18-58$.
We observe that although the general trends are reproduced, there is a visible mismatch between experimental data and predictions, 
and there is no characteristic staggering patterns 
(weakly seen for $b$ coefficients of doublets and pronounced ``sea-saw'' effect of $c$ coefficients for triplets).
For large $A$ values, one often uses an approximate form of $b$ coefficients, namely
	\begin{equation}
		b =-\frac{3 e^2}{5 r_0}A^{2/3} + \Delta_{nH} \, ,
	\end{equation}
showing that the leading order term in $b$ coefficients is proportional to $A^{2/3}$.

We do not discuss the trend of the $a$ coefficients, since it requires all other members of the liquid-drop model 
(known as the Weizs\"acker formula) and therefore a careful fitting of its parameters~\cite{BohrMott}.

We also remark that it was noticed long ago that the Coulomb interaction alone, even accurately treated, 
cannot describe observed Coulomb displacement energies of isobaric multiplets~\cite{Nolen_Schiffer69}. 
This conclusion gave rise to a special interest to nuclear models capable to
provide other than Coulomb isovector or isotensor contributions to the nuclear binding energy (e.g., see Ref.~\cite{Janecke1975} and references therein
for the earlier work on this subject). 
It is exactly the purpose of this study to address these issues within the macroscopic-microscopic approach 
of Refs.~\cite{MoellerNixADNDT1995,MoellerADNDT2016}.

\section{IMME coefficients from the macroscopic part of the FRLDM}

\subsection{General trend}

The macroscopic part of the FRLDM proposes a more involved expression for the Coulomb contribution to the atomic
mass excess as compared to the liquid-drop model.
We adopt the following formulation proposed by M\"{o}ller et al. in~Refs.\cite{MoellerNixADNDT1995,MoellerADNDT2016}:
\begin{eqnarray}
M(Z,A) & = & M_H Z + M_n (A-Z) \label{l0} \\
 & - & a_v (1-k_v I^2) A \label{l1} \\
 & + & a_s (1-k_s I^2) B_1 A^{2/3} +a_0 A^0 B_W   \label{l2} \\
 & + & c_1 Z^2B_3/A - c_4 Z^{4/3}B_s/A^{1/3}  \label{l3} \\
 & + & f(k_fr_p)Z^2/A - c_a(N-Z) \label{l4} \\
 & + & E_W + E_{\rm pair}-a_{el}Z^{2.39}\, , \label{l5} 
\end{eqnarray}
where the first two terms are mass excesses of $Z$ hydrogens and $A-Z=N$ neutrons,
followed by the volume term in line (\ref{l1}), then  
by the surface term and the so-called $A^0$ energy, a constant, that can be seen in line (\ref{l2}). 
The parameter $B_1$ defines the generalized surface energy in the model and
for a spherical nucleus can be evaluated as
\begin{equation}
	B_1=1-\frac{3}{x_0^2}+(1+x_0)\left(2+\frac{3}{x_0}+\frac{3}{x_0^3}\right)\exp{(-2x_0)} ,
\end{equation}
with $x_0=\beta A^{1/3}$, where $r_0=1.16$~fm and $\beta =r_0/a \approx 1.706$. We adopt here $B_W=1$~\cite{MoellerNixADNDT1995}.

Line (\ref{l3}) contains the direct and exchange Coulomb term with
$c_1 = 3e^2/(5r_0)$ and $c_4=5/4(3/(2\pi ))^{2/3} c_1$. 
The parameter $B_3$, defining the relative Coulomb energy
for an arbitrary shape nucleus, has in leading order (for a spherical nucleus) the following expression:
\begin{equation}
	B_3=1-\frac{5}{y_0^2}+\frac{75}{8y_0^3}-\frac{105}{8y_0^5}
\end{equation}
with $y_0=\alpha A^{1/3}$, and $\alpha =r_0/a_{den} \approx 1.657$. 
We suppose here that the relative surface energy parameter is $B_s=1$ (a spherical nucleus).

The proton form-factor correction to the Coulomb energy, $f(k_fr_p)$, is parameterized as
\begin{eqnarray}
\label{Formfactor}
	f(k_Fr_p)=&-&\frac{r_p^2 e^2}{8r_0^3}\left[\frac{145}{48} -\frac{327}{2880}(k_Fr_p)^2 \right. \\ \nonumber
	 &+& \left. \frac{1527}{1209600}(k_Fr_p)^4 \right],
\end{eqnarray}
with the proton radius, $r_p=0.8$~fm, and $k_F$ being the Fermi wave number:
\begin{equation}
\label{kF}
	k_F=\left(\frac{9 \pi Z}{4A} \right)^{1/3} \frac{1}{r_0}. 
\end{equation}
The proton form-factor depends on $A$ and $T_z$ and varies between about $-0.212$ and $-0.215$ for the nuclei of interest, with the average value being $-0.2138$~MeV for $T_z=0$ nuclei.
We have checked that the resulting rms errors for the IMME coefficients are not very sensitive to this parameter, so we kept the average value for our estimation.
The charge-asymmetry term (the second term in line (\ref{l4})), with the strength $c_a$, 
is of pure isovector character and, hence, it will contribute to the  $b$ coefficient only 
(see below Eq.~(\ref{eq:MoellerNix_b})). 

The first two terms in line~(\ref{l5}) are the Wigner contribution, $E_W$, 
\begin{equation}
\label{eq:Wigner}
E_W =W|I|+\left\{ 
\begin{array}{ll}
1/A, &  {\rm if} \; Z=N \; {\rm odd} \\
0, & {\rm otherwise}, \\
\end{array}
\right.
\end{equation}
and the average pairing term, parameterized as
\begin{equation}
\label{eq:Pairing}
E_{\rm pair} =\left\{ 
\begin{array}{ll}
\overline{\Delta}_n+ \overline{\Delta}_p - \delta_{pn}, &  {\rm if} \; Z \; {\rm odd}, \; N \; {\rm odd}\\
\overline{\Delta}_p, & {\rm if} \; Z \; {\rm odd}, \; N \; {\rm even}\\
\overline{\Delta}_n, & {\rm if} \; Z \; {\rm even}, \; N \; {\rm odd}\\
0, & {\rm if} \; Z \; {\rm even},\;  N\;  {\rm even.}\\
\end{array}
\right.
\end{equation}
The average neutron and proton pairing gaps and the average neutron-proton interaction energy 
have been parameterized as
\begin{equation}
\label{eq:NP}
\begin{array}{l}
\overline{\Delta}_n=\dfrac{r_{n}B_s}{N^{1/3}},\\[1mm] 
\overline{\Delta}_p=\dfrac{r_{p}B_s}{Z^{1/3}}, \\[1mm] 
\delta_{np}=\dfrac{h}{B_s A^{2/3}},
\end{array}
\end{equation}
with $ r_{n}=r_{p}=r_{\rm mac}=4.80$~MeV being a macroscopic pairing gap parameter. In the original
work, this parameter is kept the same for protons and neutrons.
In the present study, we adopted two sets of numerical values of various constants and parameters
from both Refs.~\cite{MoellerNixADNDT1995,MoellerADNDT2016}.
More details on the model can be found in those references as well.
Regarding the macroscopic part of the FRLDM, 
we find always a somewhat better description when using the parametrization from 
1995~\cite{MoellerADNDT2016}, so we keep this particular set of parameters in the present section.

The last term in line~(\ref{l5}) is the energy of the bound electrons with $a_{el}=1.433 \times 10^{-5}$ MeV.
Due to the smallness of this quantity we neglect this term in the present study.

It is obvious that Eq.~(\ref{l0}--\ref{l4}) can be expressed in terms of $T_z$ and $A$, 
and thus it can be cast into the form of the IMME.
Not considering for the moment the Wigner term and the pairing contribution and keeping up to quadratic 
terms in $1/A$, one can get the following (approximate) expressions for the IMME $a$, $b$ and $c$ coefficients:
\begin{eqnarray}		
a & = & \left(\frac{M_H{+}M_n}{2}-a_V-\frac{5c_1}{4 \alpha ^2} -\frac{c_4}{2^{4/3}} 
+\frac{f(k_f r_p)}{4}\right)A \nonumber\\ 
  & + & \frac{1}{4} c_1 A^{5/3}+\left(a_S+ \frac{75}{32 \alpha ^3}\right) A^{2/3} \nonumber\\ 
  & - & \frac{3 a_S}{\beta ^2}  +a_0 A^0 B_W - \frac{105}{32 \alpha^5} \, , \label{eq:MoellerNix_a} \\
b & = & \Delta_{nH}-c_1 A^{2/3}-\frac{75c_1}{8 \alpha ^3}A^{-1/3} -\frac{105 c_1}{8 \alpha ^5} A^{-1}\, \nonumber\\
		  & -&\frac{5c_1}{\alpha ^2}-\frac{4 c_4}{3(2)^{1/3}} +f(k_f r_p) +2 c_a \, , \label{eq:MoellerNix_b} \\
c & =& 4a_Vk_V+ \left(c_1-4a_Sk_S\right) A^{-1/3}\nonumber\\ 
  & -& \left(\frac{5c_1}{\alpha ^2}+\frac{4 c_4}{9(2)^{1/3}}-\frac{12a_Sk_S}{\beta^2} - f(k_f r_p) \right)A^{-1} \, \nonumber\\
		  & + & \frac{75c_1}{8 \alpha ^3}A^{-4/3} -\frac{105 c_1}{8 \alpha ^5} A^{-2}\, .
		  \label{eq:MoellerNix_c}
\end{eqnarray}
For doublets ($T=1/2$), it is straightforward to include contributions of the Wigner term and of the pairing term
to the $a$ coefficient which provides the following addition to the $a$ coefficient above:
\begin{equation}		
\Delta a _{\rm W + pair} = \frac{W}{A} - r_{\rm mac} B_S \left(\frac{2}{A}\right)^{1/3} ,
\end{equation}
while no contribution is expected to the doublets' $b$ coefficients if $r_n=r_p$.
The resulting numerical expressions are
\begin{eqnarray}
a & = & 0.1862 A^{5/3} -8.939A + 21.586 A^{2/3} \, \nonumber \\
  & - & 19.420 + 6.048 A^{-1/3}  + 30 A^{-1}\, [\textnormal{MeV}] \,, \label{eq:MoellerNix_a_num} \\
b & = & -0.7448 A^{2/3}  + 2.748 - 1.534 A^{-1/3} \, \nonumber \\
	 & + & 0.7823 A^{-1} \, [\textnormal{MeV}] \label{eq:MoellerNix_b_num}\,.
\end{eqnarray}
These relations represent a very good approximation to the theoretical $a$ and $b$ coefficients for doublets obtained from the full macroscopic part of FRLDM and are shown in Fig.~(a,b) (labeled as FL-MAC) in comparison with the available experimental data. 
First, we observe that the general trends of $a$ and $b$ coefficients are well reproduced.
For $a$ coefficients, it is interesting to note that FRLDM predicts that the Coulomb term takes over beyond $A\approx 75$, 
where the $a$ coefficient curve reaches its minimum, and starts to grow for larger values of $A$. 
To asses the quality of the results, we use here the rms error defined as
\begin{equation}
\label{rmse}
{\rm rmse}(x) = \sqrt{ \frac1n\sum_{i = 1}^{n} \left[ x_{th}(i)-x_{exp}(i) \right]^2}\,,
\end{equation}
where $x$ stands for $a$, $b$ or other quantity, $x_{th}(i)$ are theoretical values, $x_{exp}(i)$ are experimental values
and $i$ runs over the available data points ($i = 1,\ldots, n$). 
The resulting rms errors of the macroscopic part of the FRLDM for $a$ and $b$ coefficients can be found in Table~I 
(see the entry FL-MAC). 

The original parametrization results in relatively large rms errors for the IMME coefficients
as seen from Table I.
This is due to the fact that the parameters have been fixed together with the microscopic part (shell correction) of the model.
The rms deviations can be greatly improved, if we adjust some selected parameters.
To this end, we choose three parameters, namely, $a_0$, $c_a$ and $r_{\rm mac}$ which contribute to the isoscalar,
isovector and isotensor channels, and we show that by slightly changing their values we can imrpove the description
of the $a$, $b$ and $c$ coefficients, respectively. 
The corresponding modified version is named FL-MAC-M, where ``M'' stands for ``modified''.
For example, increasing $a_0$ from 2.165 MeV to 4.7 MeV, we can reduce the rms error for 
the $a$ coefficients from 3 MeV to about 1.9 MeV. 
The modifications of $c_a$ and $r_{\rm mac}$ which influence $b$ and $c$ coefficients, respectively, will be described below.
(To be precise, the parameter $r_{\rm mac}$ also slightly impacts the isoscalar channel --- our modification reduces the rms error on 
the $a$ coefficient by additional 30 keV).
The results for the $a$ coefficients obtained from FL-MAC-M are shown in Fig. 2(a). 

\begin{figure}
 \centering
  \includegraphics[width=0.45\textwidth]{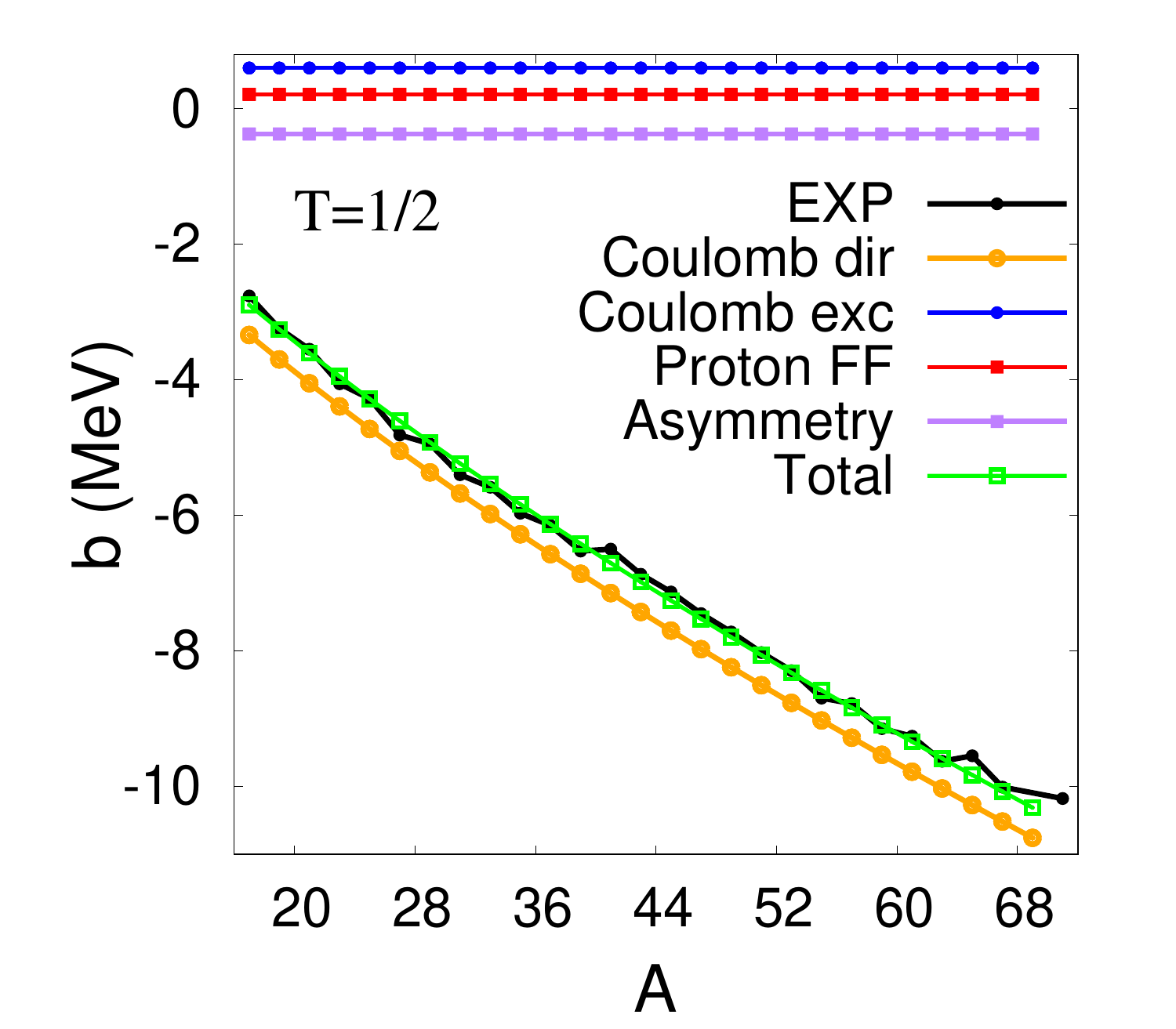}
  \caption{\label{fig:b-contrib} (Color online) Theoretical IMME $b$ coefficients for doublets: the total value  (from FL-MAC-M)
and the contribution of different terms from Eq.~(\protect\ref{eq:MoellerNix_b}) in comparison with experimental data. 
The asymmetry parameter is set to $c_a=0.186$~MeV. See text for further details.} 
\end{figure}

The macroscopic part of the FRLDM predicts rather well both the trend and the magnitude of $b$ coefficients,
leading to a much better agreement with the data than the liquid-drop model. 
However, it can be noticed that the $b$ coefficients from the FRLDM have a kind of offset with respect to the data.
To address this issue, we study various contributions to the $b$ coefficients as easily seen from 
Eq.~(\ref{eq:MoellerNix_b}): the Coulomb direct and exchange terms,
the proton form-factor contribution and the asymmetry term. 
The general trend is assured by the term ($\sim A^{2/3}$), dominating the Coulomb contribution. 
The other terms provide (in our approximation) only small constant shifts of $b$ coefficients.
Among them, we notice that the proton-neutron asymmetry term contributes only to the $b$ coefficient. 
We have varied this term and found that a global improvement in the rms error for $b$ coefficients can be achieved by slightly changing its value from $c_a=0.10289$~MeV to $c_a=0.186$~MeV: 
the rms error reduces for a total number of data points reduces from 189~keV to 91~keV 
(see the entry FL-MAC-M in Table~I).
Fig.~\ref{fig:b-contrib} shows this improvement and illustrates the role of various terms to the $b$ coefficient value, 
with the new value of $c_a$ taken into account. 
From now, we therefore keep this optimized value of $c_a$ in our modified version of the macroscopic part of the FRLDM.

For $c$ coefficients, the implication of Eq.~(\ref{l0}--\ref{l4}) is not well defined. 
As we mentioned in Section~II, the FRLDM is designed to describe nuclear masses, 
while to get the $c$ coefficient we have to know the excitation energies of $|T_z|<T$ isobaric-multiplet members 
(e.g., of the $T_z=0$ member in triplets).
Following Ref.~\cite{LamPRC2013}, we can restrict ourselves by assuming that only Coulomb terms (both direct and exchange ones) 
contribute to the $c$ coefficient. This means that we suppose $k_V=k_S=0$ in Eq.~(\ref{eq:MoellerNix_c}), 
which results in the following numerical expression:
\begin{eqnarray}
\label{eq:MoellerNix_c_num}
c & = & 0.7448 A^{-1/3}- 1.771 A^{-1} + 1.535 A^{-4/3} \, \nonumber \\
  & - & 0.7823 A^{-2} \, [\textnormal{MeV}] \,. 
\end{eqnarray}
The corresponding overall trend of $c$ coefficients is shown in  Fig.~\ref{fig:abc-trend}(c) 
(a smooth curve named FL-MAC-C) in comparison with the experimental data on $c$ coefficients for triplets and 
the liquid-drop model (LDM) predictions.

Alternatively, we can use all isotensor terms of the macroscopic part of the FRLDM, as well as the contribution
from the Wigner and pairing terms. Bearing in mind, that the model approximates
nuclear masses, we take into account the experimentally
observed excitation energies of the IAS in $T_z=0$ nuclei in order to reconstruct isobaric triplets.
By doing this, indeed we get an overall description including a staggering effect.
The original parametrization results in the rms error for the $c$ coefficients of triplets of around 841~keV.
Again, we checked that increasing a value of the average pairing gap parameter $r_{\rm mac}$  from 4.80~MeV to 5.51~MeV, 
we can reduce the rms deviation to 632~keV. 
This value is still too big as compared to the absolute value of the $c$ coefficients (between about 200 and 350~keV), as is easily
seen from the resulting curve for the $c$ coefficients shown in Fig. 2(d) (labeled as FL-MAC-M).
Although the rms error is reduced, we observe, however, that theoretical $c$ coefficients exhibit staggering of a very large amplitude
as compared to experiment. Only up to $A=28$, this staggering is in phase with the data. 
For heavier nuclei, oscillations become irregular.
This means that the isotensor part of the model is not well constrained by the fit and, therefore, 
not suitable for understanding the $c$ coefficients.
In Section V, we will see that the addition of the microscopic part does not remedy the situation.

While it looks difficult to refine the description of the IMME $c$ coefficients in detail,
we would be interested in getting a staggering pattern of the $b$ coefficients. 
This is however not provided by the formal analytical expressions of the original macroscopic part of the FRLDM. 
We address this question in the next section, carefully considering the pairing energy parameterization.

\subsection{Staggering pattern of $b$ coefficients and average proton and neutron pairing gaps}

The $r_{\rm mac}$ parameter of the macroscopic part of the FRLDM is an average between proton and neutron
pairing energies. As is explained in the original work~\cite{MoellerNixADNDT1995,MoellerADNDT2016},
its magnitude is not important for the full model, 
since it is the microscopic part which is added and optimized to ensure a good agreement with experiment.
However, a precise magnitude of the average pairing in the macroscopic part can help us to understand the staggering phenomenon.
In the context of different models, this effect has also been discussed in  Refs.~\cite{Janecke1966a,LamPRC2013,Kaneko2013}.

\begin{figure*}
 \centering
  \includegraphics[width=0.3\textwidth]{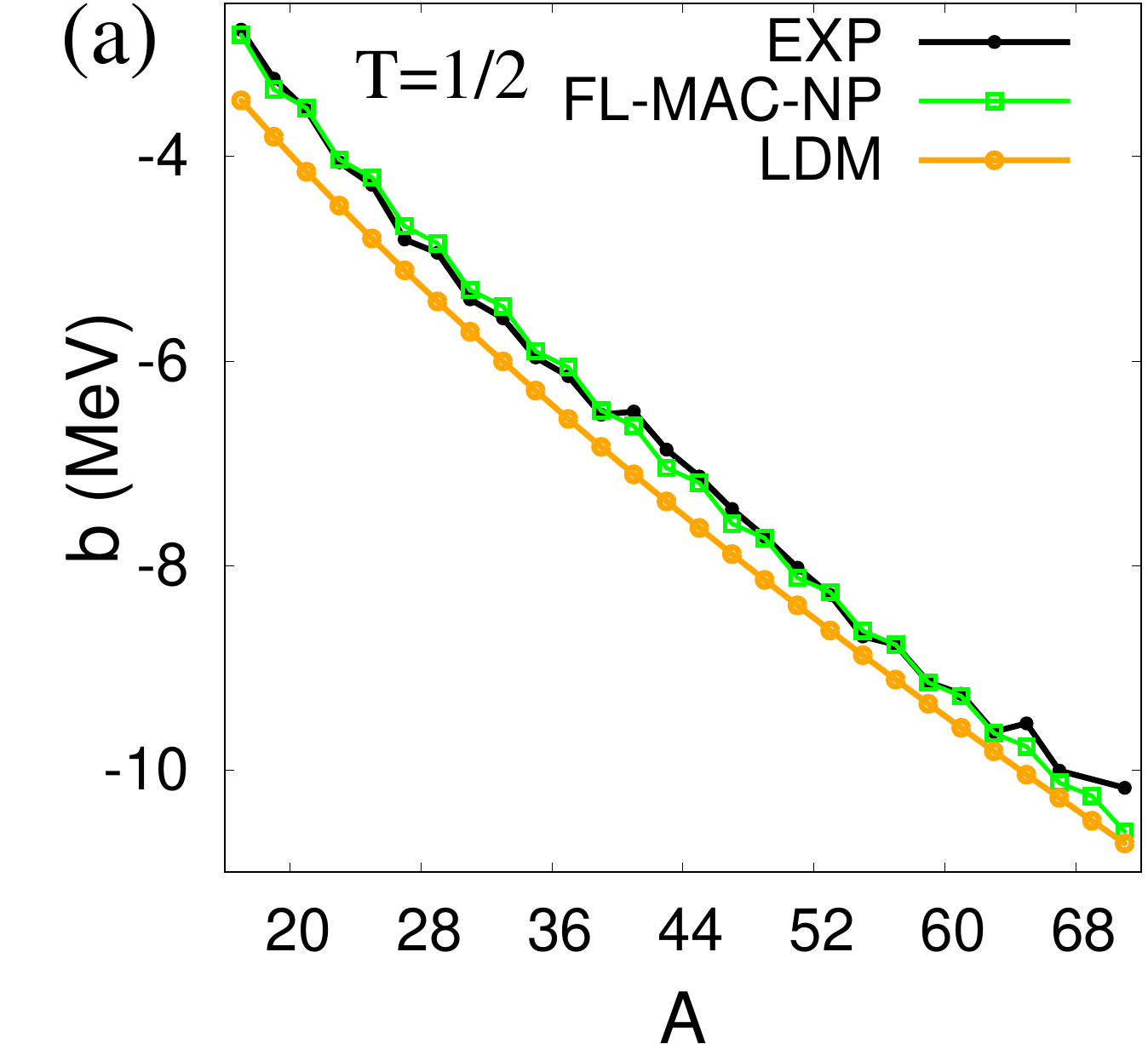} \hspace{3mm}
  \includegraphics[width=0.3\textwidth]{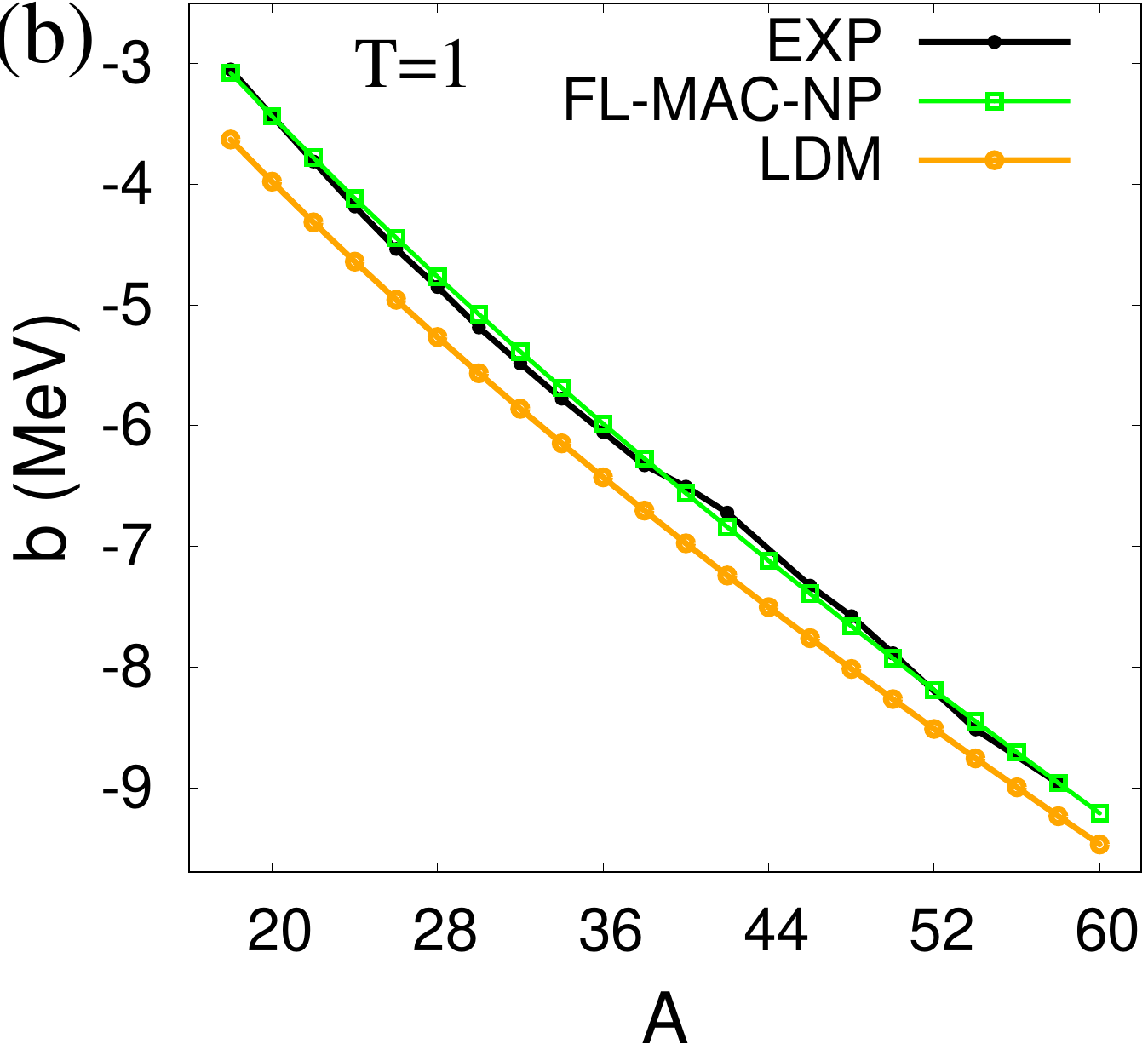} \hspace{3mm}
  \includegraphics[width=0.3\textwidth]{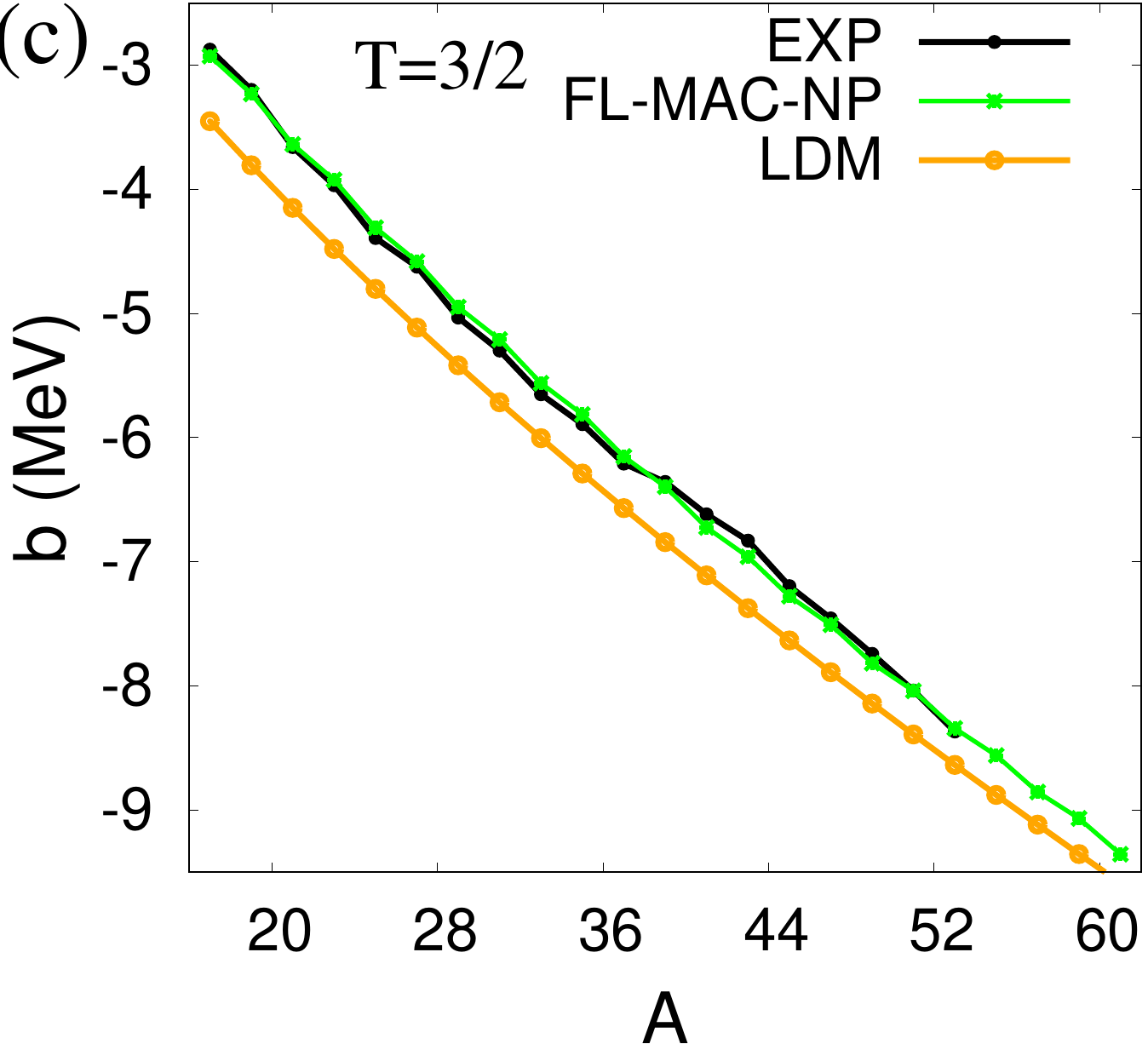}
  \caption{\label{fig:b_np}  (Color online) Theoretical IMME $b$ coefficients for doublets (a), triplets (b) and quartets (c)
obtained from the macroscopic part of the FRDLM with different proton and neutron $r_n$ and $r_p$ parameters 
(FL-MAC-NP) in comparison with experiment.} 
\end{figure*}

\begin{figure*}
 \centering
  \includegraphics[width=0.19\textwidth]{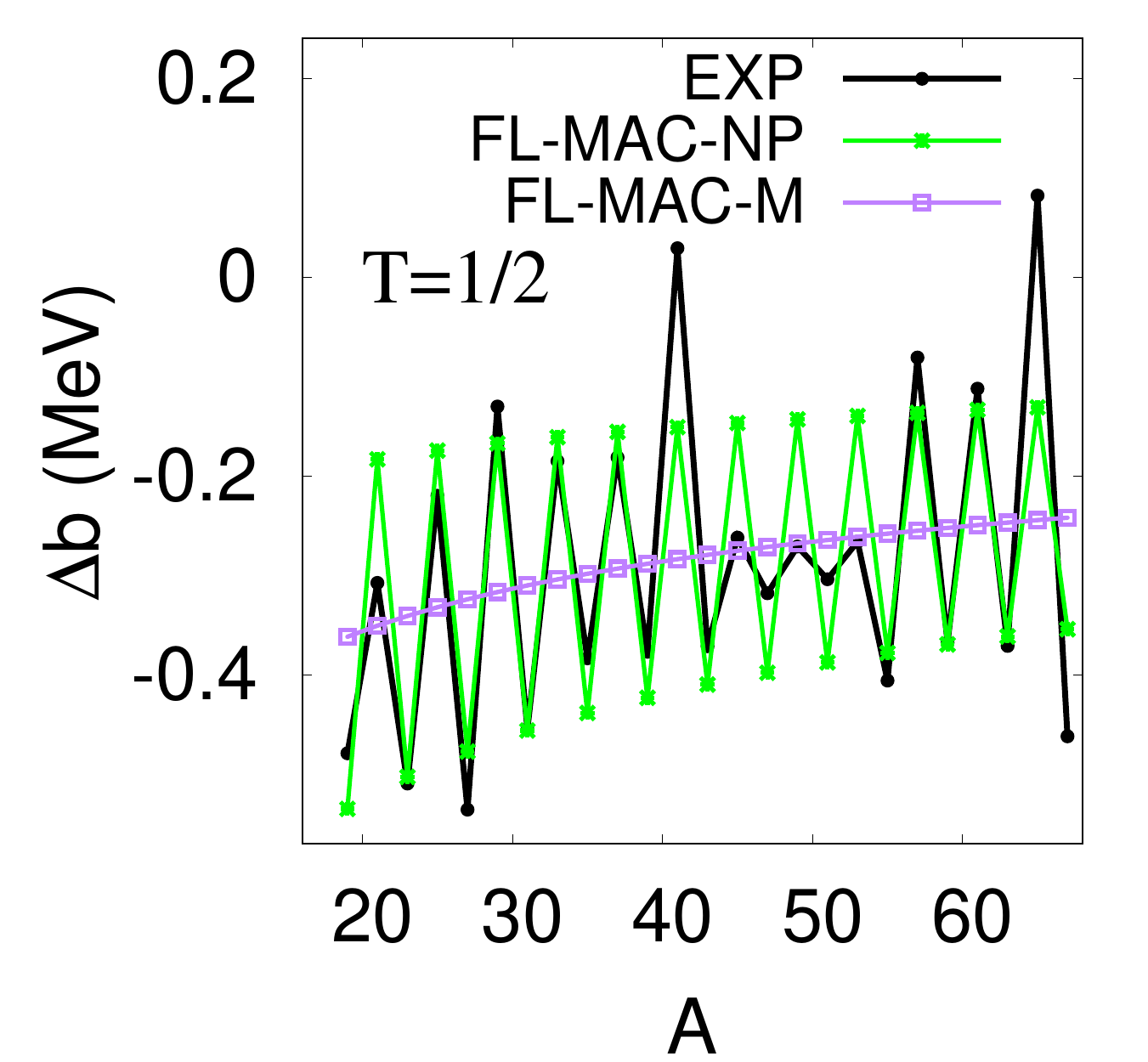}  
  \includegraphics[width=0.19\textwidth]{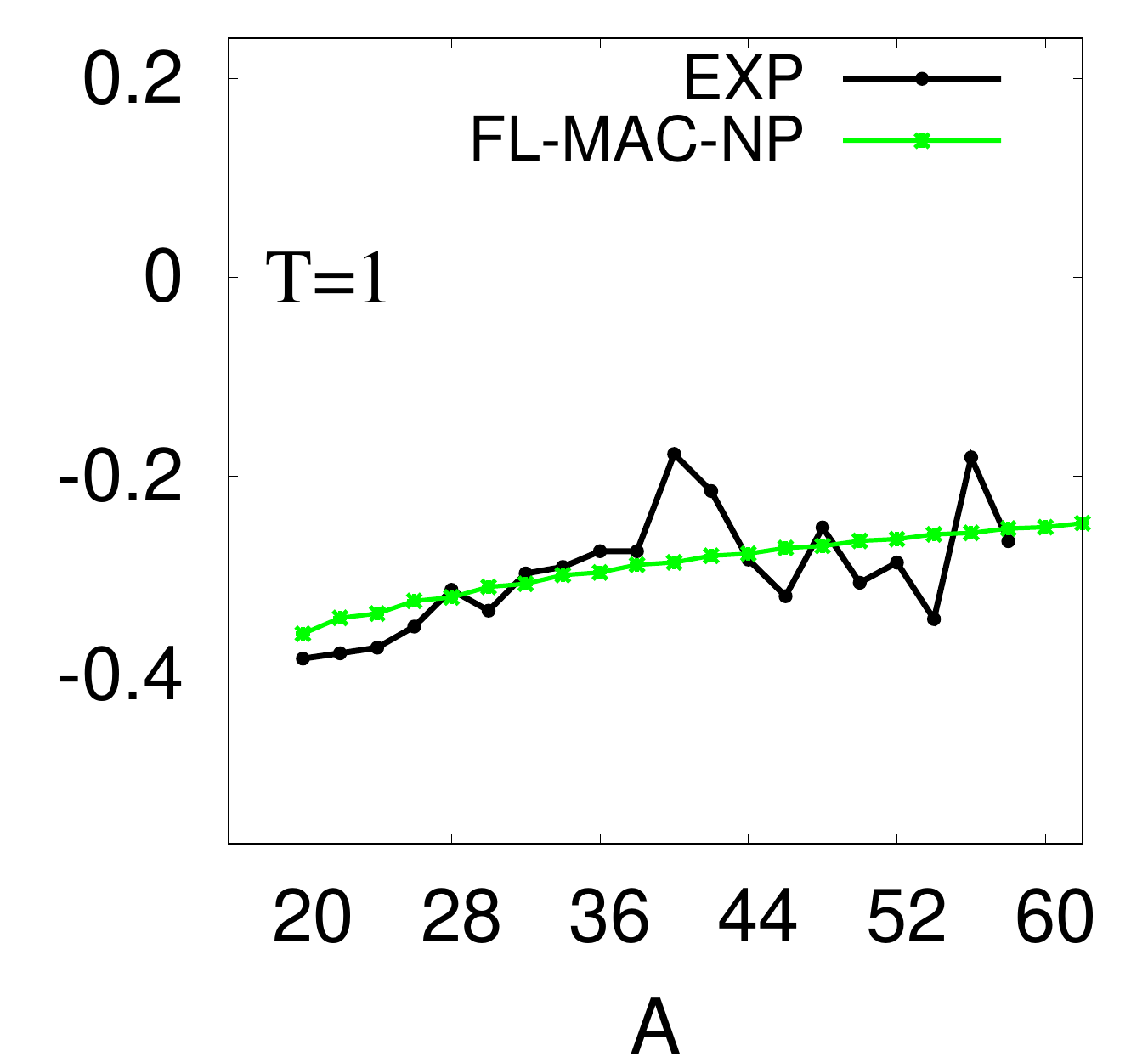}  
  \includegraphics[width=0.19\textwidth]{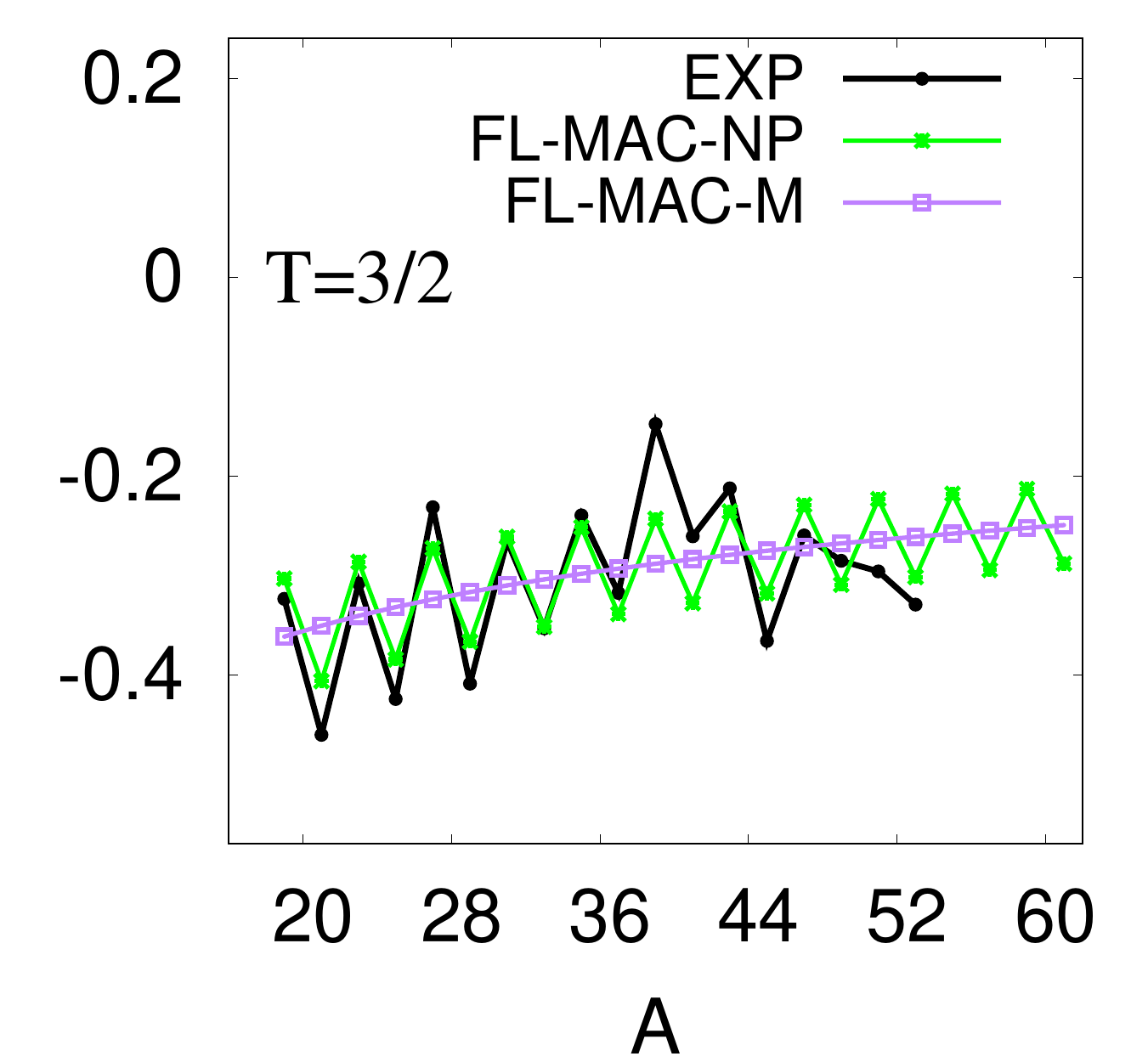} 
    \includegraphics[width=0.19\textwidth]{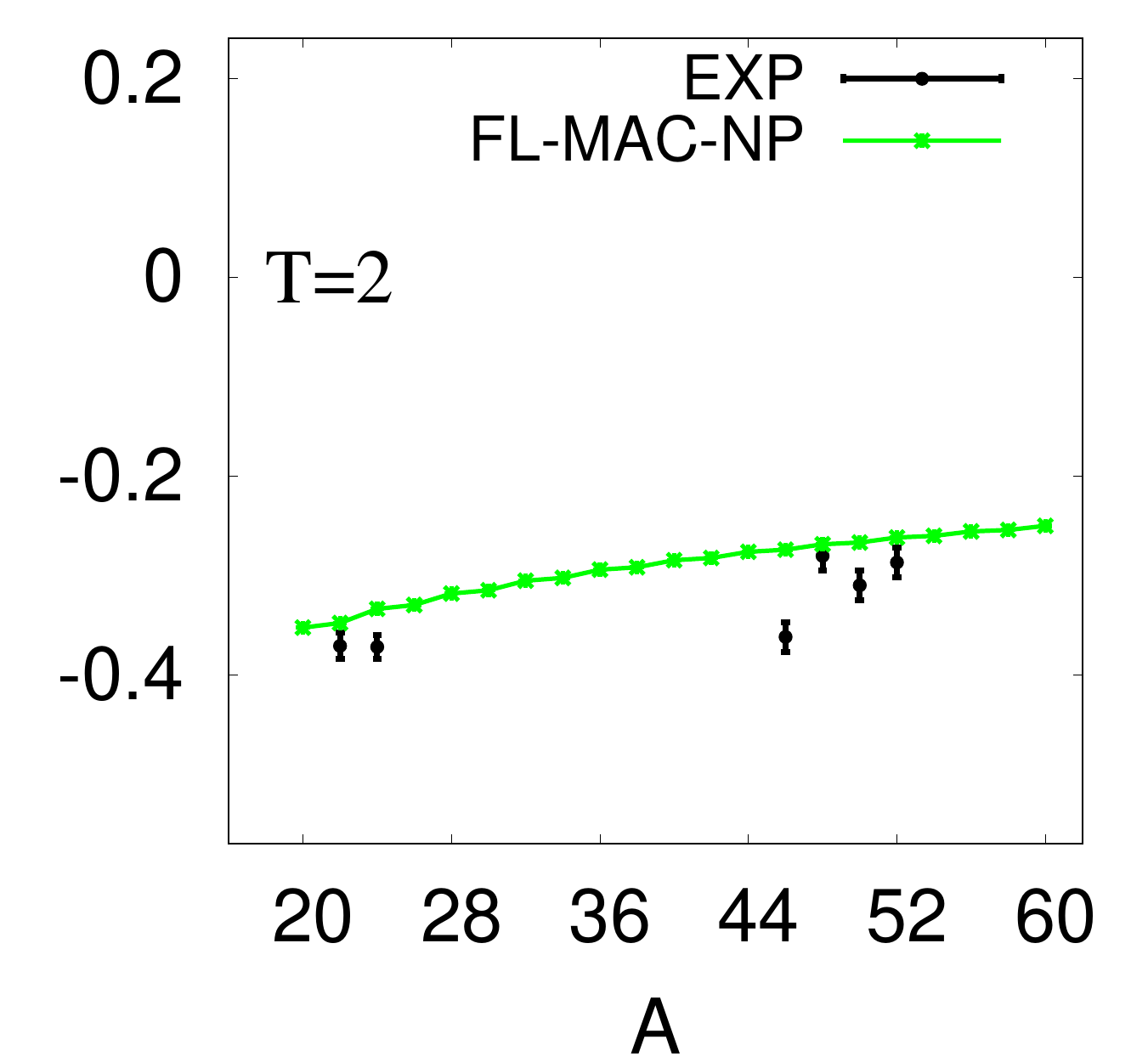} 
  \includegraphics[width=0.19\textwidth]{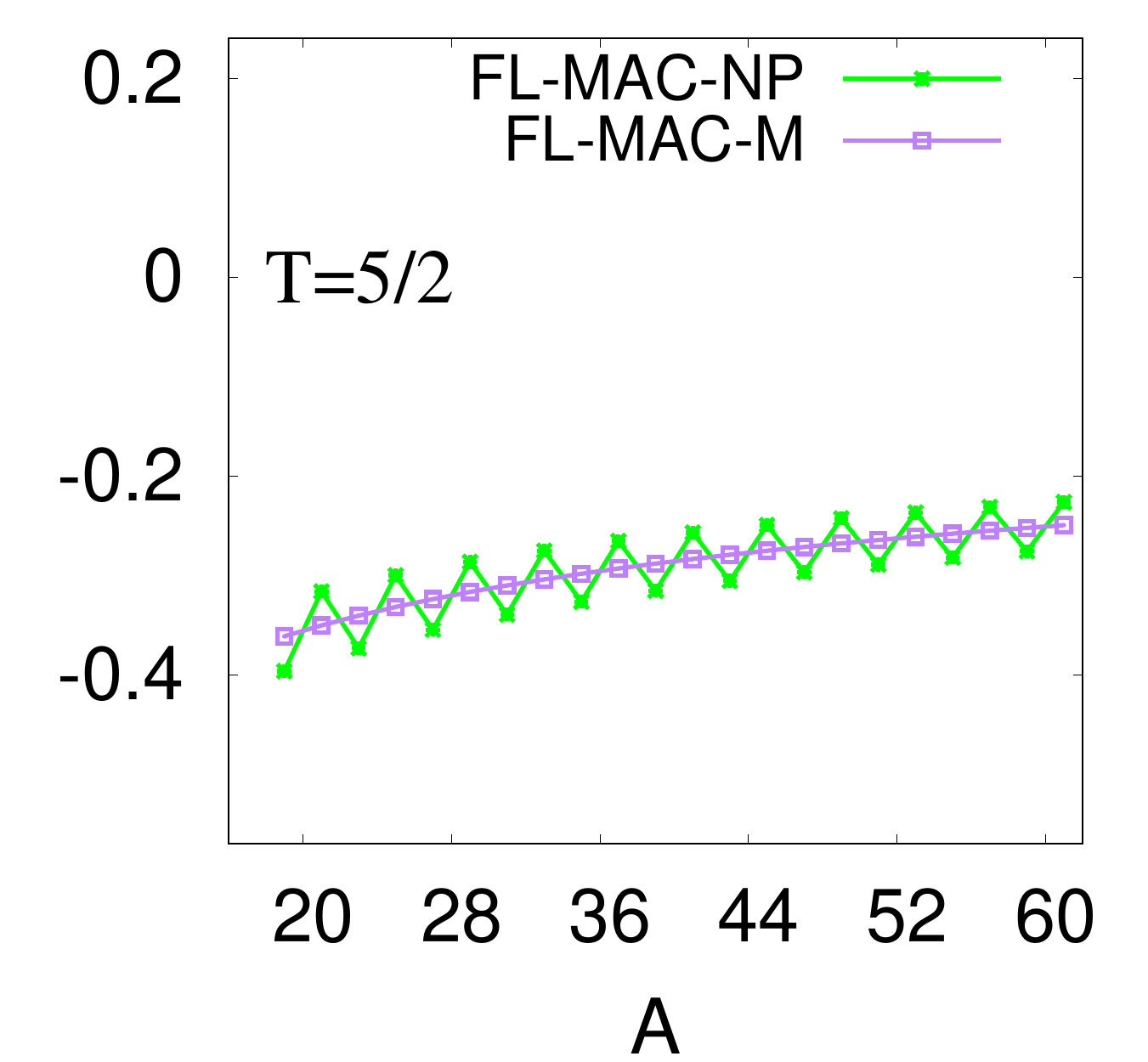} 
  \caption{\label{fig:staggering}  (Color online) Differences of IMME $b$ coefficients, $\Delta b =b(A)-b(A-2)$, 
for $T=1/2,1,3/2,2,5/2$ multiplets as obtained from the modified macroscopic part of the FRDLM with either identical proton and neutron pairing gap parameters (FL-MAC-M), or with different proton and neutron pairing gap parameters (FL-MAC-NP), in comparison with experiment. 
See text for details.} 
\end{figure*}

From the study of proton and neutron pairing gaps, we know that those proton and neutron energies may be different. 
Following the work of M\"oller and Nix~\citep{MoellerNix1992}, to determine the $r_n$ and $r_p$ parameters 
separately, we performed a least-squares fit of $\overline{\Delta}_n$ and $\overline{\Delta}_n$ from Eqs.~(\ref{eq:NP})
to the experimental neutron and proton pairing gaps, respetively, of nuclei along the $N=Z$ line from $A=12$ to $A=74$, 
which are of interest in our work.
The established values are $r_n=6.83$~MeV and $r_p=6.65$~MeV, resulting in a difference of
$r_n-r_p = 0.018$~MeV. 
It is this difference between $r_n$ and $r_p$ that leads to the staggering of the $b$ coefficients, as we will show below. 
The obtained rms errors are summarized in the third line of Table~I, referred to as FL-MAC-NP 
(the parameter $c_a=0.186$~MeV, as was proposed in the previous section).
We see that there is a further improvement in the rms error for the $b$ coefficients compared to FL-MAC-M: 
it reduces in total from 91~keV to 81~keV.
The corresponding $b$ coefficients for $T=1/2, 1, 3/2$ multiplets are shown in comparison with experimental data in Fig.~\ref{fig:b_np}.
Indeed, we notice staggering of the $b$ coefficients for doublets and triplets which is in accord with the experimental data. 
To better see this effect, we plot in Fig.~\ref{fig:staggering} the quantity $\Delta b =b(A)-b(A-2)$ for various $T$-multiplets 
in comparison with the available experimental data.

The results on staggering can be understood analytically. The contribution of the pairing term to  $b$ coefficients of $A=4n+1$ and $A=4n+3$ doublets are
\begin{equation}
\label{eq:bpair12}
b_{pair}^{T=1/2}=\left\{
\begin{array}{l}
(r_n-r_p) \dfrac{2^{1/3} B_s}{(A+1)^{1/3}}, \quad A=4n+1\\
(r_p-r_n) \dfrac{2^{1/3} B_s}{(A-1)^{1/3}}, \quad A=4n+3
\end{array}
\right.
\end{equation}
which results in a staggering amplitude $\Delta b=b(A)-b(A-2)$ of
\begin{equation}
\label{eq:b_stag12}
\Delta b^{T=1/2}\approx 2(r_n-r_p) \dfrac{2^{1/3} B_s}{A^{1/3}},
\end{equation}
if we assume $A=4n+1$.
At the same time, for quartets we get the following contributions to the $b$ coefficients:
\begin{equation}
\label{eq:bpair32}
b_{pair}^{T=3/2}=\left\{
\begin{array}{l}
(r_p-r_n) \dfrac{2^{1/3} B_s}{3(A+3)^{1/3}}, \quad A=4n+1\\
(r_n-r_p) \dfrac{2^{1/3} B_s}{3(A-3)^{1/3}}, \quad A=4n+3
\end{array}
\right.
\end{equation}
which results in a staggering amplitude of
\begin{equation}
\label{eq:b_stag32}
\Delta b^{T=3/2}\approx -2(r_n-r_p) \dfrac{2^{1/3} B_s}{3A^{1/3}},
\end{equation}
if we assume $A=4n+1$.
This is about of $1/3$ of the amplitude of staggering for doublets in absolute value and is out of phase.
The resulting trends of $\Delta b$ can be seen in Fig.~\ref{fig:staggering}  for $T=1/2$ on the first panel and for $T=3/2$ multiplets on the third panel in comparison with available experimental data. 
For reference, we also plot $\Delta b$ for half-integer values of $T=1/2, 3/2$ and $5/2$ that 
comes from the modified macroscopic part of the FRLDM with $r_n=r_p$ (FL-MAC-M): it does not show any staggering. 

Estimation of the staggering effect for $b$ coefficients of triplets shows that 
\begin{equation}
\label{eq:bpair1}
b_{pair}^{T=1}=\left\{
\begin{array}{l}
0, \quad A=4n+2\\
2(r_p-r_n) \dfrac{2^{1/3} B_s}{3A^{4/3}}, \quad A=4n
\end{array}
\right.
\end{equation}
which results in a staggering amplitude  $\Delta b=b(A)-b(A-2)$ of
\begin{equation}
\label{eq:b_stag1}
\Delta b^{T=1}\approx 2(r_n-r_p) \dfrac{2^{1/3} B_s}{3A^{4/3}},
\end{equation}
if we assume that $A=4n+2$.
This is a factor of $1/(3A)$ and $1/A$ smaller as compared to the staggering in $b$ coefficients for doublets or quartets, respectively
(the effect cancels in the leading order of the Taylor expansion). 
As we can observe from Fig.~\ref{fig:staggering}, second panel, the staggering is indeed hardly visible on the scale of the figure, while the general trend is in robust agreement with experiment.
The value of $\Delta b$, obtained from FL-MAC-M is not shown in the plots for $T=1$ (and $T=2$), 
because it coincides with the present theoretical curve FL-MAC-NP in the scale of the figure.

Similarly, going further, we get that the staggering effect for $b$ coefficients of quintets ($T=2$) is negligible, as given by the analytical expression
\begin{equation}
\label{eq:b_stag2}
\Delta b^{T=2}\approx -2(r_n-r_p) \dfrac{2^{1/3} B_s}{3A^{4/3}},
\end{equation}
assuming that $A=4n+2$. These are of the similar magnitude, but out of phase compared to those for triplets.
For $T=5/2$ multiplets, the staggering is again stronger, and in phase with the $T=1/2$ pattern,
being five times smaller in amplitude, according to 
\begin{equation}
\label{eq:b_stag52}
\Delta b^{T=5/2}\approx 2(r_n-r_p) \dfrac{2^{1/3} B_s}{5A^{1/3}}.
\end{equation}
The resulting trends can be seen in Fig.~\ref{fig:staggering} (last two panels).
These patterns are expected to hold for higher half-integer and integer $T$ multiplets. 
Future experiments will verify these predictions. 

We remark that although the macroscopic part of the FRLDM can grasp the overall trend and phases of this peculiar staggering behavior,
it can predict only a smooth variation of the staggering amplitude with mass because of the ansatz given by Eq.~(\ref{eq:NP}).
This is due to the fact that the pairing effect is accounted in an average way.
At the same time, we observe variations of the staggering amplitude in Fig.~5 which are probably related to the shell effects and 
the underlying microscopic structure of the many-nucleon systems. 
For example, for doublets, we notice stronger variations in $b$ value for $A=41$ and $A=65, 67$ and oscillations of
very small amplitude in the $f_{7/2}$ nuclei between $A=43$ and $A=53$ (Fig. 5, left panel).
Similarly, various irregularities can be noticed for other multiplets.
The macroscopic part of the FRLDM alone cannot account for these fine structures. 
It is an accurately added shell correction or a fully microscopic model which has to deal with these effects.

\begin{table*} 
\vspace{-1ex}
\caption{\label{tab:rms} The rms errors of the IMME $a$, $b$ and $c$ coefficients calculated within the macroscopic part of the FRLDM (denoted as FL-MAC), as well as from the full versions of the FRDM and FRLDM as given in Ref.\protect\cite{MoellerNixADNDT1995} (1995) and \protect\cite{MoellerADNDT2016} (2016). Calculations labeled as FL-MAC-M refer to updated values of $a_0$, $c_a$ and $r_{mac}$ with respect to the original model, 
while  FL-MAC-NP exploits in addition the different neutron and proton pairing gap parameters as explained in Section IV.B. 
See text for details.}
\begin{ruledtabular}
\begin{tabular}{l|c|c|c|c|c|c}
Model & \multicolumn{2}{c|}{$T=1/2$ ($A$=17--67) } & \multicolumn{2}{c|}{$T=1$ ($A$=18--58)} & $T=3/2$ ($A$=19--39) & Total \\
\hline
      & rmse $a$ (keV) & rmse $b$ (keV) & rmse $b$ (keV) & rmse $c$ (keV) & rmse $b$ (keV) & rmse $b$ (keV) \\
\hline
FL-MAC            & 3009  & 189  & 191  & 841 & 186 & 189 \\[1mm] 
FL-MAC-M          & 1857  & 113  &  72  & 632 &  75 &  91 \\[1mm] 
FL-MAC-NP         & 1857  &  95  &  71  & 632 &  70 &  81 \\[1mm] 
FRLDM (1995)      & 1210  & 454  & 416  & 984 & 442 & 438 \\[1mm] 
FRLDM (2016)      & 1109  & 434  & 469  & 484 & 529 & 467 \\[1mm] 
FRDM (1995)       & 1145  & 321  & 250  & 823 & 231 & 278 \\[1mm] 
FRDM (2016)       & 1062  & 246  & 234 & 1038 & 227 & 238 \\[1mm] 
\end{tabular}
\end{ruledtabular}
\end{table*} 

\subsection{IMME beyond the second order}

The mass excess parameterization proposed by the FRLDM allows to extend the IMME beyond the second order in $T_z$.
This can be done by expanding in Taylor series the contribution of the Coulomb exchange term up to the fourth order
in $T_z$. The extended IMME thus would be of the form:
\begin{eqnarray}
M(\eta ,T,T_z) & = & a(\eta ,T) + b(\eta ,T) T_z + c(\eta ,T) T_z^2 \nonumber \\
               & + & d(\eta ,T) T_z^3+ e(\eta ,T) T_z^4\,, \label{eq:IMMEext}
\end{eqnarray}
where $d$ and $e$ coefficients come out to have the following expressions:
\begin{equation}
\label{eq:de}
d =  \dfrac{4 \, c_4}{2^{1/3} \, 81 \, A^2},\ \quad
e = -\dfrac{40 \, c_4}{2^{1/3} \, 243 \, A^3}\,.
\end{equation}
Indeed, these quantities appear to be very small, the $d$ coefficients are positive, while the $e$ coefficients are always negative. 
The numerical values are larger for lower masses.
For example, for $A=17$,  $d\approx 77$~eV and $e\approx -15$~eV, while for $A=8$, these values would be about 350~eV and $-150$~eV,
respectively. These values are considerably smaller than those few non-zero cases determined experimentally (which are all of the order
of keV, e.g. Refs.~\cite{ADNDT2013,MacCormick2014}). 

There might be a specific contribution from the pairing term as well as from the proton form-factor,
but these are expected to be of even smaller magnitude.

\section{IMME coefficients from the FRLDM and FRDM}

\begin{figure*}
 \centering
  \includegraphics[width=0.28\textwidth]{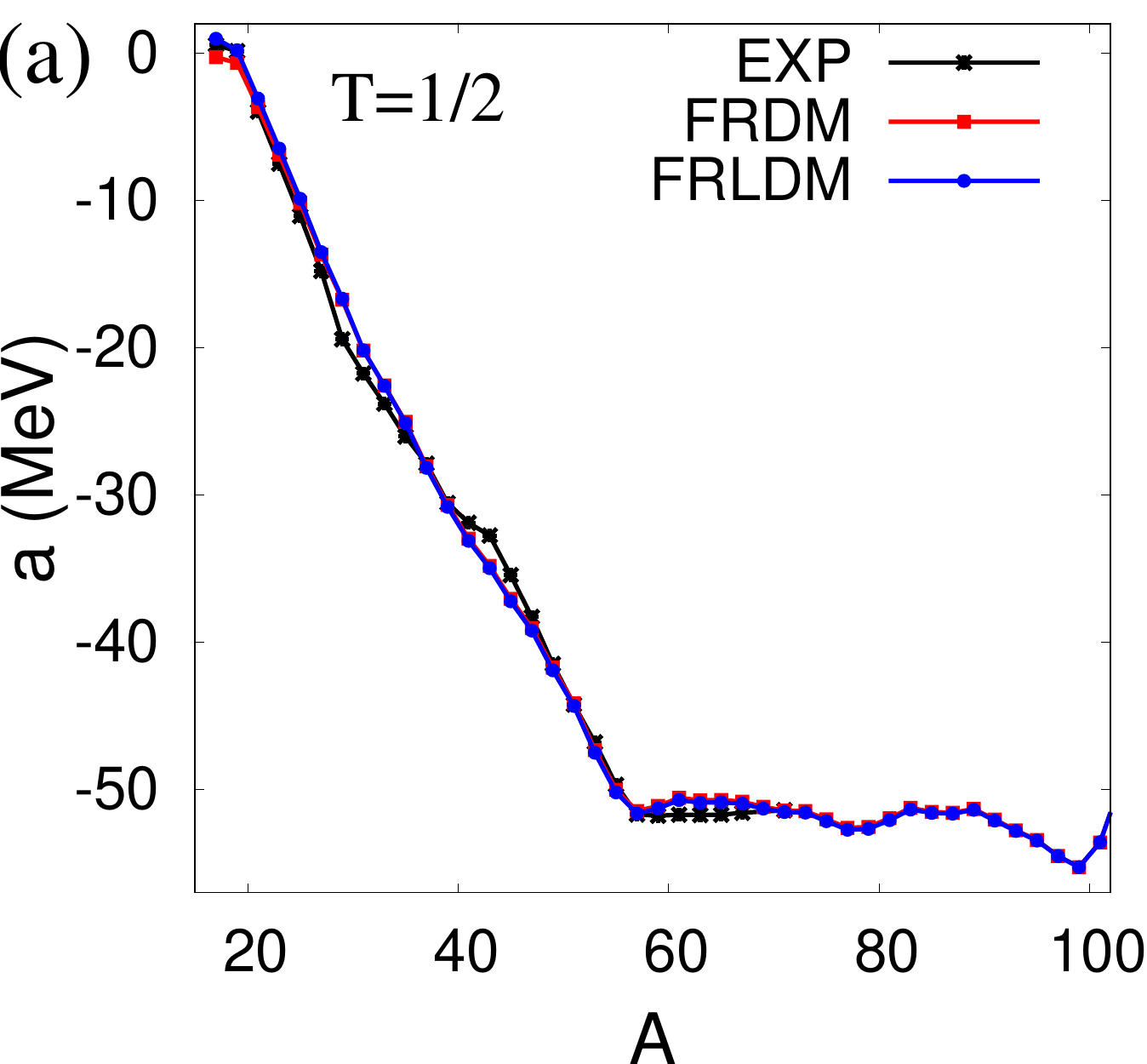} \hspace{2mm}
  \includegraphics[width=0.28\textwidth]{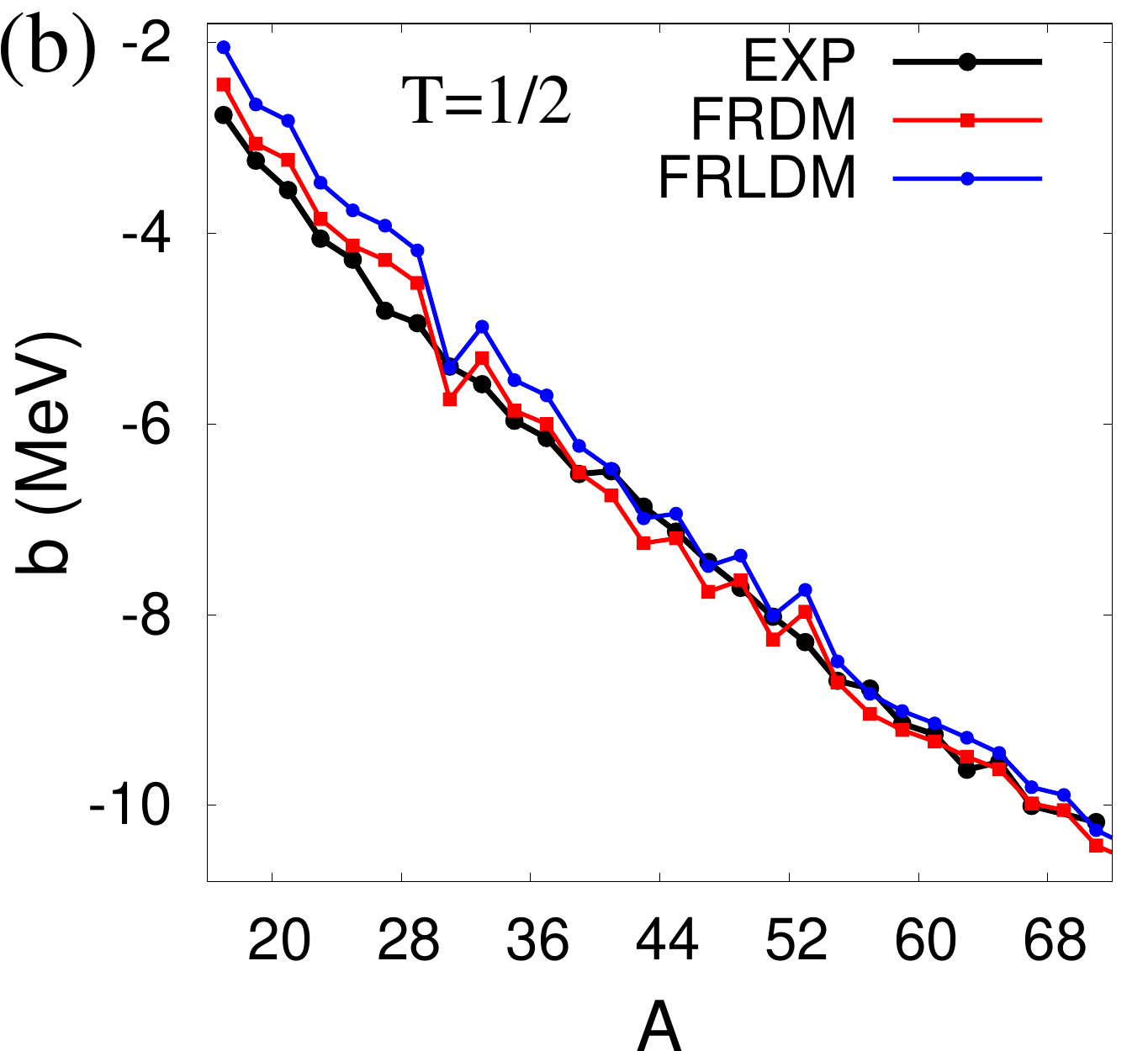} \hspace{2mm}
  \includegraphics[width=0.28\textwidth]{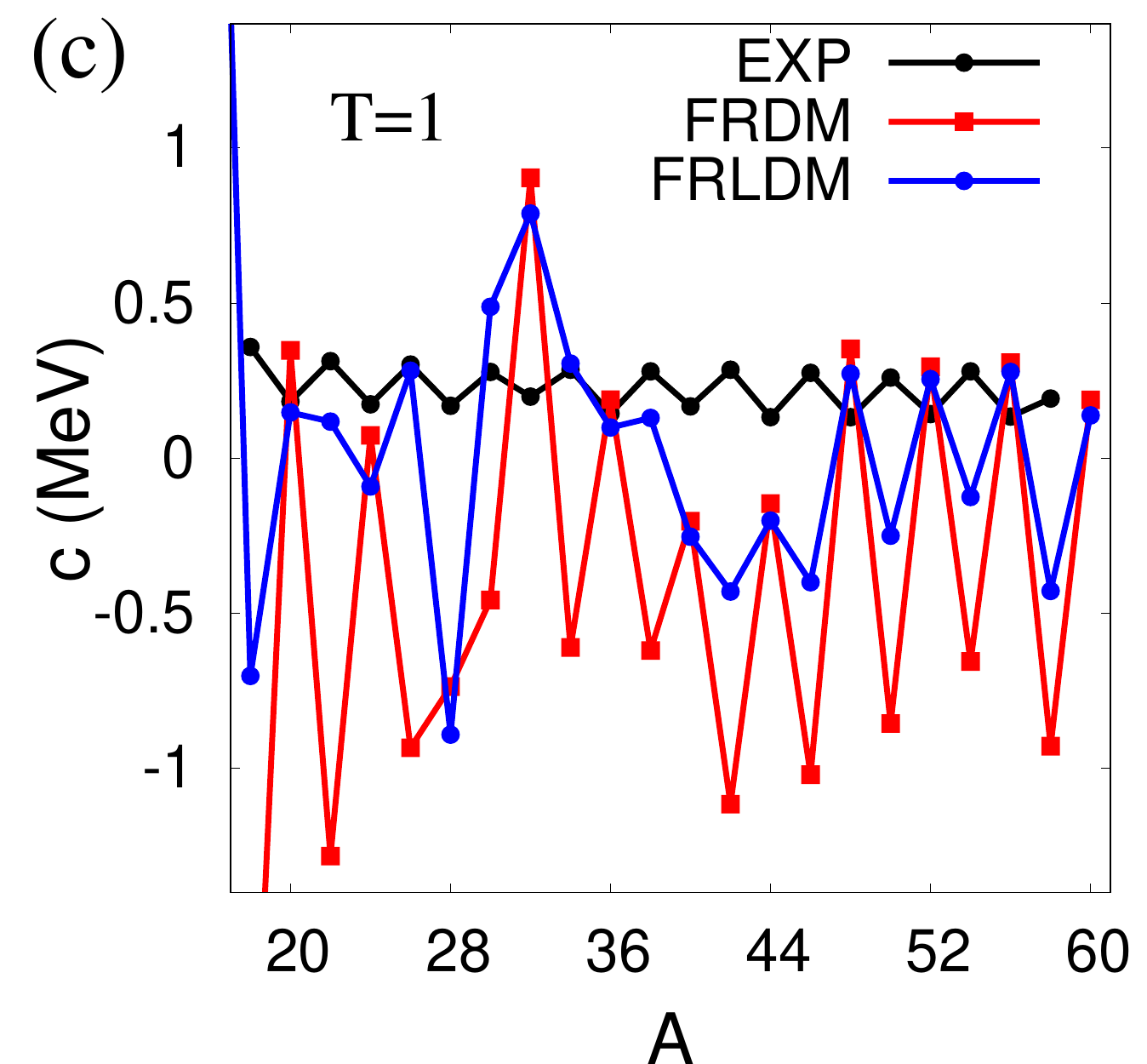} 
  \caption{\label{fig:ab_FRDM}  (Color online) Theoretical IMME $a$ and $b$ coefficients (panels (a) and (b), respectively)
 and $c$ coefficients for triplets (c) obtained from the full FRDM and FRLDM
 in comparison with the experimental data. 
 To deduce theoretical $c$ coefficients we used theoretical masses from the FRDM/FRLDM and
  experimental excitation energies in $T_z=0$ nuclei.} 
\end{figure*}

In this section, we deduce the IMME $a$, $b$ and $c$ coefficients using the mass predictions
from the full FRDM and FRLDM for multiplets with $A>16$ up to 101 for $a$ coefficients, up to $A=71$ for $b$ coefficients
and up to $A=60$ for $c$ coefficients.
The results obtained from the values of the mass compilation~\cite{MoellerADNDT2016} are shown in Fig.~\ref{fig:ab_FRDM}, 
while the rms errors are summarized in Table I for both mass compilations,
the one from 1995 \cite{MoellerNixADNDT1995} and the other from 2016 \cite{MoellerADNDT2016}.
The overall rms errors are generally smaller for the latest set, with a few exceptions.

First, it is very well seen that both models, the FRDM and FRLDM, perfectly describe the $a$ coefficients for doublets, reproducing the end of the strong down slopping trend up to $A\approx 57$. 
The predicted $a$ coefficients for heavier nuclei along the $N=Z$ line is expected to stay roughly flat up to about $A\approx 90$ and 
to decrease in their absolute value towards $A=100$. 
No data exist yet to compare with.

\begin{figure*}
 \centering
  \includegraphics[width=0.3\textwidth]{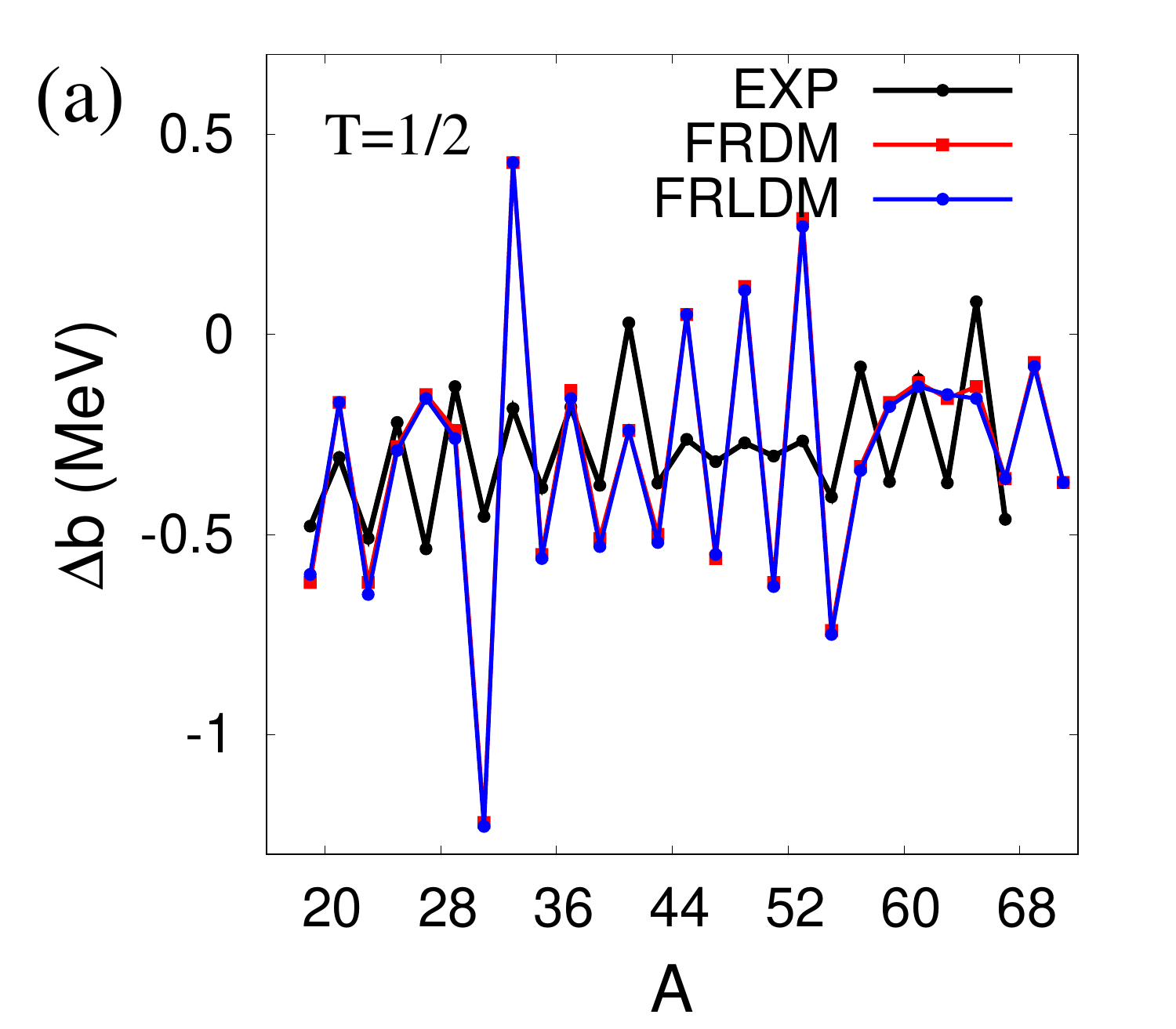}  
  \includegraphics[width=0.3\textwidth]{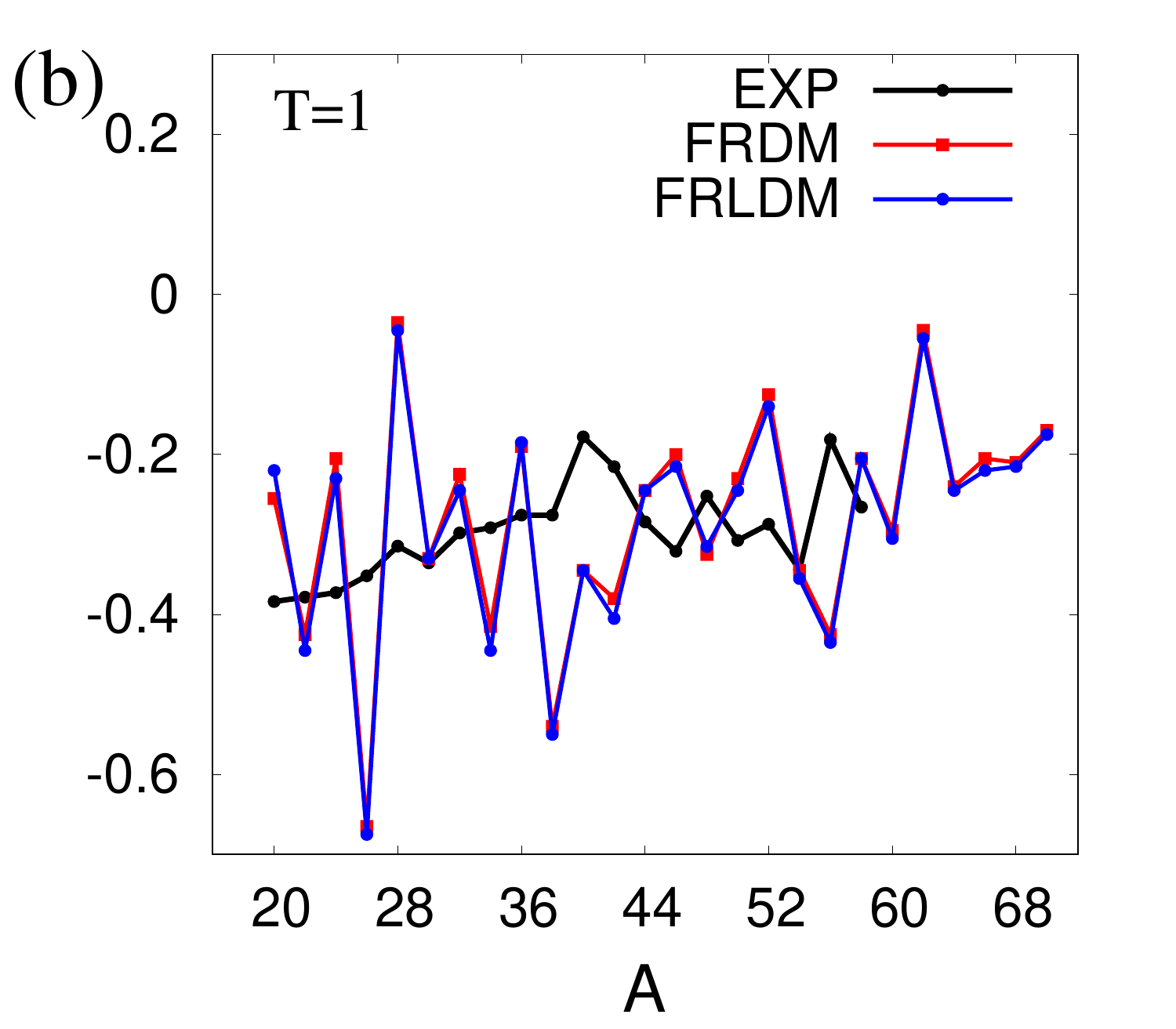}  
  \includegraphics[width=0.3\textwidth]{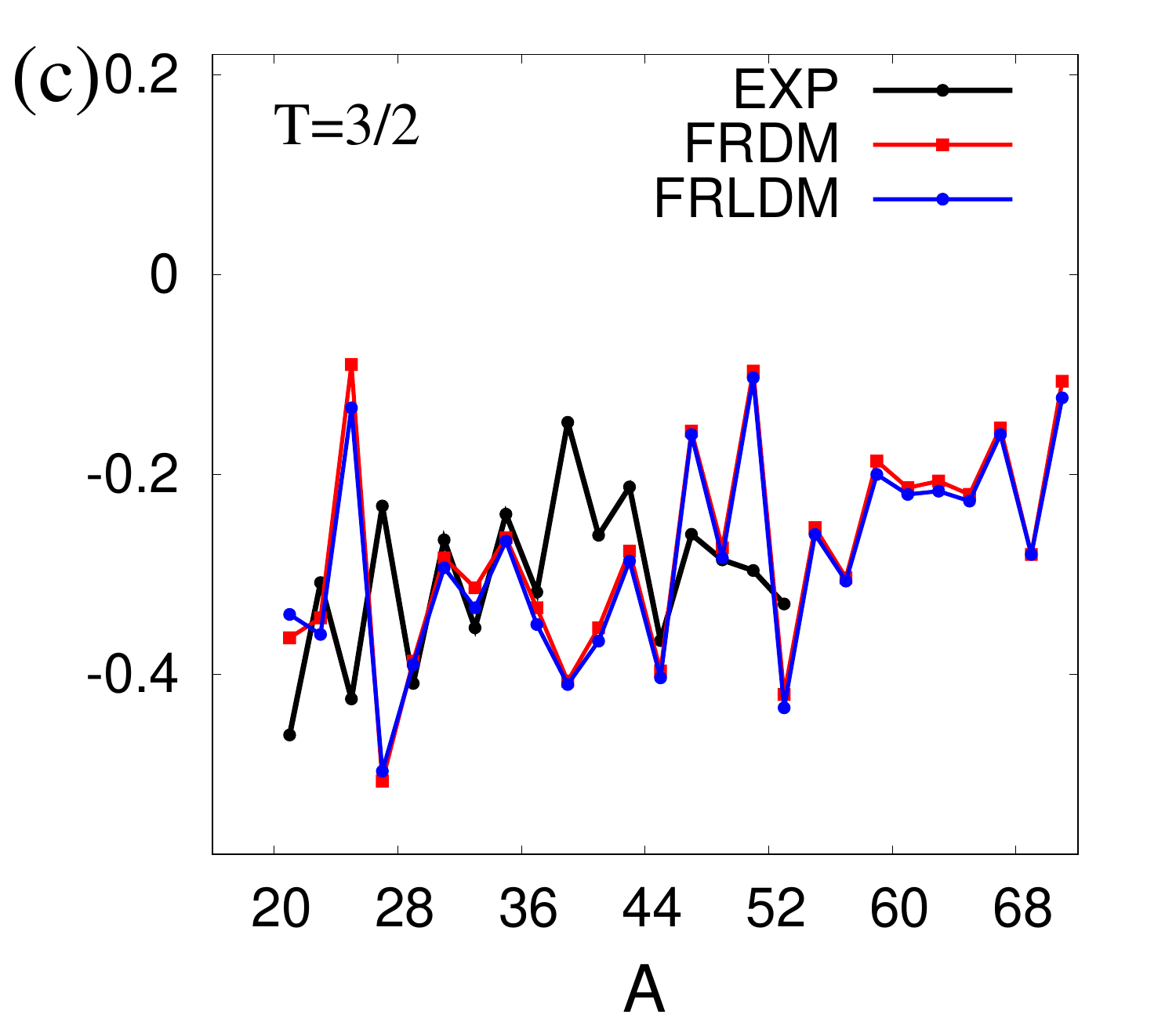} 
  \caption{\label{fig:staggering_FRDM}  (Color online) Differences of IMME $b$ coefficients, 
  $\Delta b =b(A)-b(A-2)$,
 for doublets (a), triplets (b) and quartets (c) as obtained from the FRDM and FRDLM  in comparison with the experimental data.} 
\end{figure*}

For $b$ coefficients, the agreement between the two models (FRDM and FRLDM) and the data is less convincing compared to what we could obtain from the macroscopic part of the FRLDM only (c.f. with Fig.~\ref{fig:b_np}(a)). 
The corresponding rms errors, shown in Table~I, support this conclusion.
The staggering effect is plotted in Fig.~\ref{fig:staggering_FRDM} for doublets, triplets and quartets and is seen to be too much exaggerated compared to the data and even not always in phase with experiment.
Similar conclusion is valid for $c$ coefficients, which are shown in Fig.~\ref{fig:ab_FRDM}(c): 
the precision of the FRLDM/FRDM is not sufficient to describe the data in detail.
The corresponding rms errors can be found in Table I.
This conclusion is not surprising: the overall rms error of the models is 550~keV over almost the full mass chart (almost 2150 masses).
Since $b$ and $c$ coefficients are linear combinations of two or three masses, the overall error may reach even higher values.
This prevents the detail description.
We also believe that considering the experimental constraints on the $b$ and $c$ coefficients may help to improve on the modelization.


\section{Masses of proton-rich nuclei up to $A=100$}


One of the major impacts of nuclear masses for heavy proton-rich nuclei is related to their importance for astrophysics,
in particular, for type-I X-ray burst models~\cite{Schatz1998}. 
Special attention from experiment and theory has been focused on the so-called waiting points of the $rp$ process, 
e.g. $^{64}$Ge, $^{68}$Se and $^{72}$Kr~\cite{SchatzPRL2001}.
The proton capture on these nuclei results in a
proton-unbound or weakly bound nucleus and thus, the $\beta ^+$-decay may take over and deviate 
the $rp$-process path. 

\begin{figure*}
  \includegraphics[width = 1.1\textwidth]{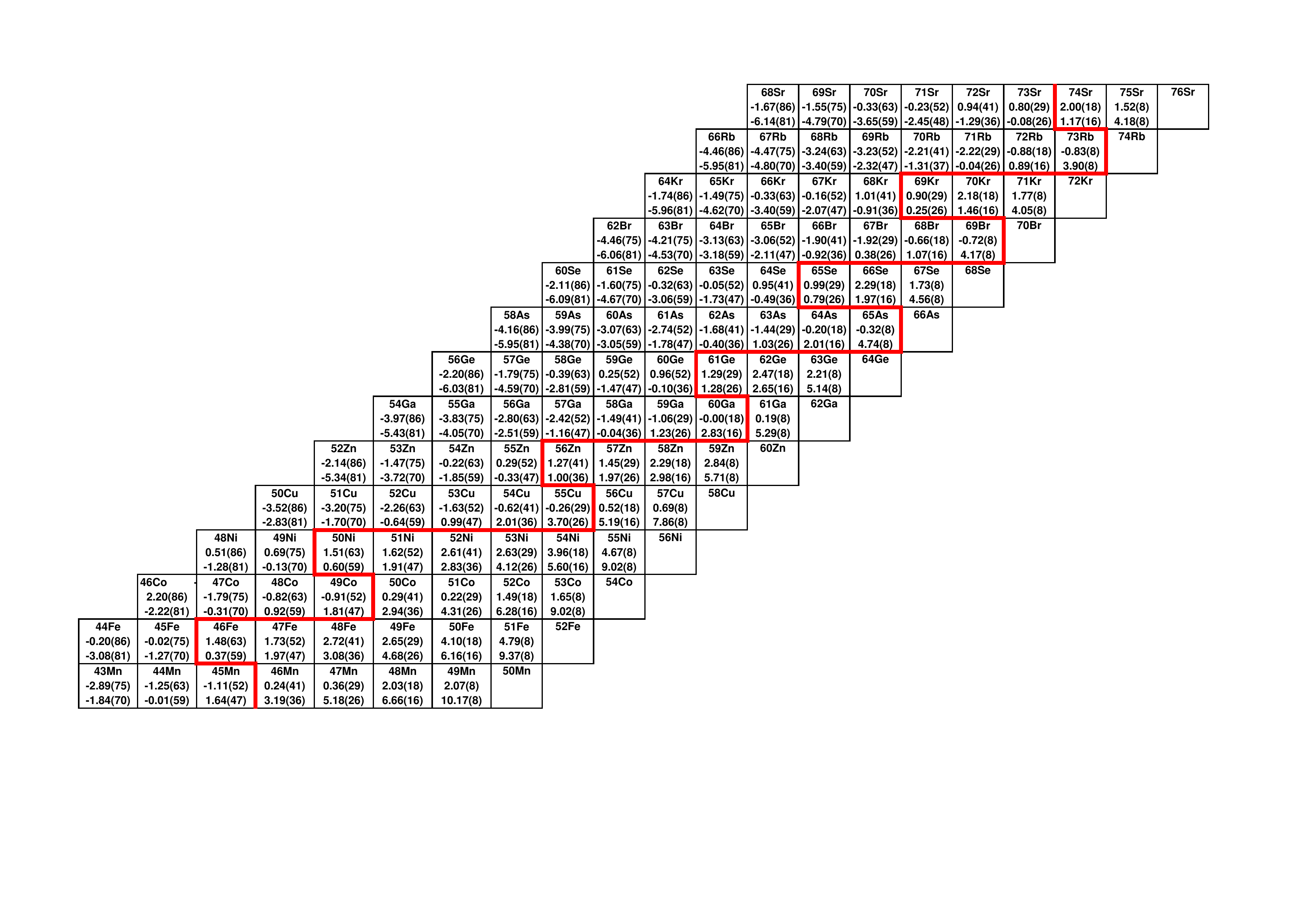} 
  \caption{\label{fig:chart1}  (Color online) A fragment of the nuclear chart for $Z>N$ nuclei with $25\le Z \le 38$.
Theoretical values of $S_p$ and $S_{2p}$ are indicated in each box below a nucleus symbol. The uncertainties
are indicated in parenthesis. 
The thick red line shows a tentative position of the proton drip-line based on the average values of $S_p$ and $S_{2p}$.
See text for details.} 
\end{figure*}

\begin{figure*}
  \hspace{-1cm}
  \includegraphics[width = 1.1\textwidth]{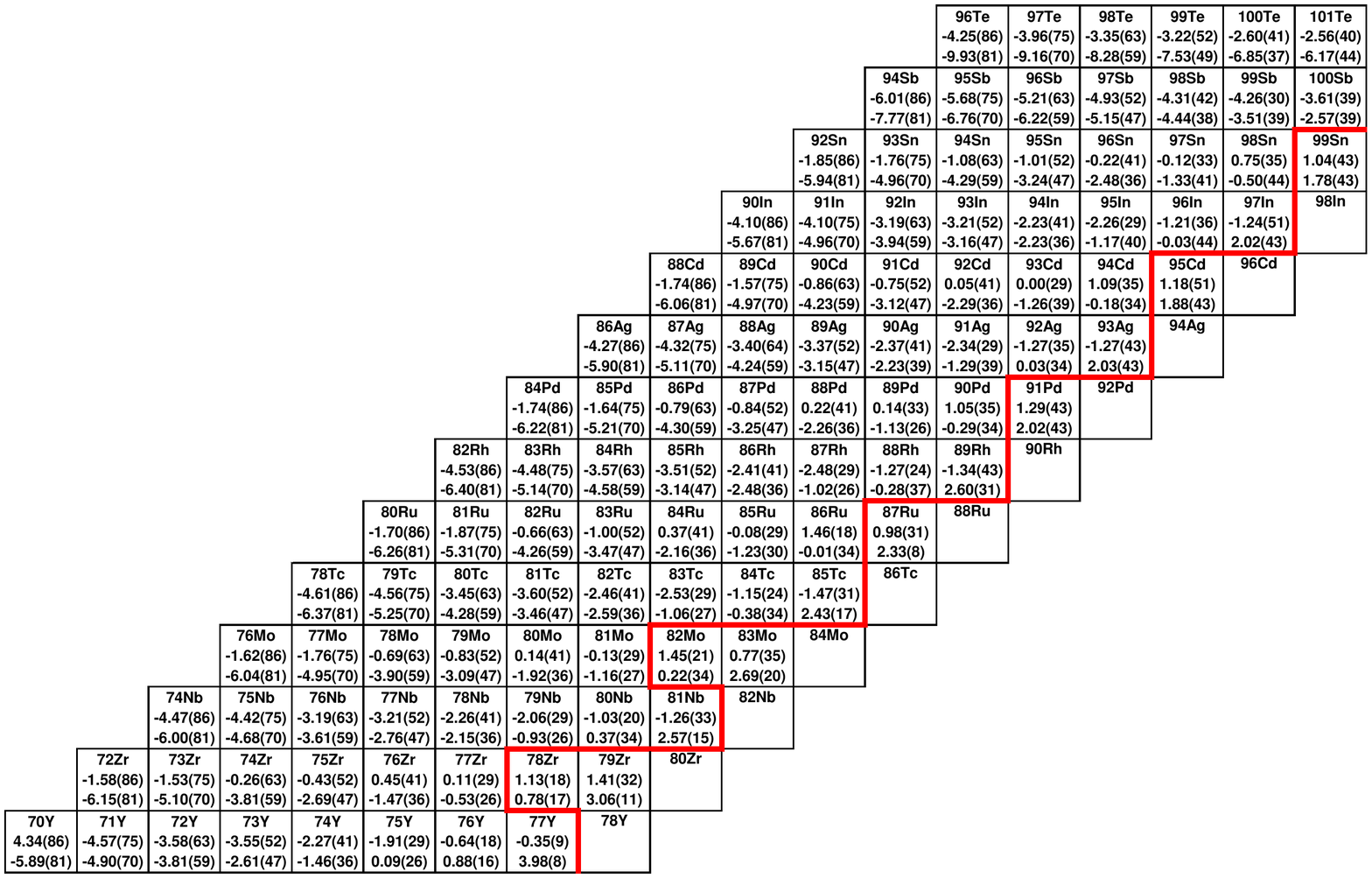} 
  \caption{\label{fig:chart2}  (Color online) A fragment of the nuclear chart for $Z>N$ nuclei with $39\le Z \le 52$.
Theoretical values of $S_p$ and $S_{2p}$ are indicated in each box below the nucleus symbol. The uncertainties
are indicated in parenthesis. 
The thick red line shows a tentative position of the proton drip-line based on the average values of $S_p$ and $S_{2p}$.
See text for details.} 
\end{figure*}

As nuclear mass increases along the $N=Z$ line, nuclei become more and more 
short lived due to increasing Coulomb effects 
and, as a result, it becomes harder and harder to discover them and determine their disintegration
modes.  Lots of experimental efforts have been devoted to explore the properties of very proton-rich nuclei
and to trace the position of the proton drip-line 
(e.g., Refs.~\cite{BlankPRL1995,BlankPRL2005,Stolz2005,RogersPRL2011,Faestermann2002,DelSanto2014,Blank2016,CelikovicPRL2016,Suzuki2017,Sinclair2019}).
Still, beyond Zr nuclei, astrophysical simulations rely only on atomic mass extrapolations~\cite{AME2016} 
or on various theoretical predictions~\cite{Brown91,Cole96,Ormand97,Brown2002,RMF2001,Olsen2013,Kaneko2013,Brown2019}.

Since our best $b$ coefficients obtained from the modified macroscopic part of the FRLDM with different
proton and neutron pairing gaps (FL-MAC-NP) results in a relatively small rms error of known $b$ coefficients of about 80~keV, 
we decided to use this model to calculate the masses of the proton-rich nuclei and thus to map the proton drip-line.
To reach this goal, we exploit the method of Coulomb displacement energies.
Using the notations from Section II, $M(\eta, T, T_z)\equiv M_{T_z}$ for a given isotopic multiplet $(\eta ,T)$,
we can express the mass excess of a proton-rich nucleus (with $T_z=-T)$ 
on the basis of an experimental mass excess of its neutron-rich mirror 
(with $T_z=T)$ and a theoretical $b$ coefficient~\cite{PaAn1988,Brown91,Ormand97,Brown2002} as follows:
\begin{equation}
\label{CDE}
M_{-T} =  M_T - 2\, b\, T \,.
\end{equation}
Theoretical $b$ coefficients have been calculated using the proton-neutron version of the modified macroscopic part of the FRLDM
(FL-MAC-NP).
The experimental masses of neutron-rich nuclei and $N=Z$ nuclei have been taken from Ref.~\cite{AME2016}.
Thus, the mass excesses of nuclei with $Z>N$ have been obtained via Eq.~(\ref{CDE}).
The resulting one- and two-proton separation energies ($S_p$ and $S_{2p}$) 
for proton-rich nuclei with $25 \le Z \le 52$ and $-4 \le T_z \le -1/2$ 
are summarized in the form of two nuclear chart fragments shown in Figs.~\ref{fig:chart1} and \ref{fig:chart2}. 
The position of the one-proton or two-proton drip-line is sketched by the thick read line, 
based on the central values of $S_p$ for odd-$Z$ isotopes or $S_{2p}$ for even-$Z$ isotopes, respectively.

The uncertainties on the obtained values of mass excesses of nuclei with $Z>N$ are estimated as
\begin{equation}
\label{unc}
\Delta(M^{\rm th}_{-T}) = \sqrt{ [\Delta(M_T^{\rm exp})]^2 + 4\, T^2 \, [\Delta(b_{\rm th})]^2},
\end{equation}
where $\Delta(b_{\rm th})\approx 81$~keV, estimated from the average rms error of theoretical $b$ coefficients, 
while  $\Delta(M_T^{\rm exp})$ refers to the uncertainty on the experimental 
mass excess of neutron-rich mirrors. It follows that for the majority of cases,
the uncertainty on the mass of a proton-rich nucleus, and therefore on the one- and two-proton separation energies,
is mainly governed by the magnitude of the $2T\Delta(b_{\rm th})$ term and, thus, is particularly large
for high-$T$ isobaric multiplets.
Nevertheless, with increasing $N$ and $Z$, the proton drip-line approaches the $N=Z$ line and our results
bring more accurate predictions for the stability of the nuclides of interest.

In general, we conclude that in spite of large uncertainties, our model provides a relatively robust picture
of this exotic region of the nuclear chart and the position of the proton (and two-proton) drip-line agrees for a majority of
cases with the predictions of Ref.~\cite{Ormand97,Brown2002,Kaneko2013}, taking into account the assigned theoretical uncertainties. 
Although we traced the drip-line on the basis of the average values of $S_p$ and $S_{2p}$,
often the uncertainty may allow to shift it one nucleus left or right.

The nuclear chart fragments in Figs.~\ref{fig:chart1} and \ref{fig:chart2} contain in total 220 theoretical 
$S_p$ and $S_{2p}$ values. 
From these numbers, only 103 sets of $S_p$ and $S_{2p}$ can be calculated based on the masses present in the AME2016 data base~\cite{AME2016}
(24 experimentally measured masses and 79 extrapolated values).
Overall, our calculated theoretical one- and two-proton separation energies
are in fair agreement with these data.
To address the quality of the results, we have calculated the rms error according to Eq.~(\ref{rmse})  
with $x$ being either $S_p$ or $S_{2p}$.
We find that ${\rm rmse}(S_p)$=320~keV and ${\rm rmse}(S_{2p})$=353~keV for a total number of $n=103$ data points.

We note, however, that extrapolated masses in AME2016~\cite{AME2016} may contain
big associated uncertainties and therefore the application of Eq.~(\ref{rmse}) may not be an optimal choice.
If we exclude all extrapolated masses in Ref.~\cite{AME2016} and estimate the rms error  for the 24 experimentally measured cases,
we get ${\rm rmse}(S_p)$=148~keV and  ${\rm rmse}(S_{2p})$=154~keV.
In addition, we have calculated the $\chi^2$ per number of degrees of freedom ($\chi^2/\nu $)  with $\nu = n-1$,
\begin{equation}
\label{chi2}
\chi^2/\nu = \frac{1}{n-1} \sum_{i = 1}^{n} \dfrac{\left[ S_{th}(i)-S_{exp}(i) \right]^2}{\sigma_{th}^2(i) + \sigma_{exp}^2(i)}\,,
\end{equation}
and get that for $n=24$ experimentally measured cases, $\chi^2/\nu =0.88$ for $S_p$, while $\chi^2/\nu =0.94$ for $S_{2p}$.
These close to the unity estimates show that the theoretical outcome is reasonable.


Now, let us briefly discuss the results.
In the lower part of the nuclear chart (Fig.~\ref{fig:chart1}), the position of the drip-line 
and the properties of near lying nuclei are largely known experimentally~\cite{BlankBorge08,PfutznerRMP2012}.
According to our calculation, the odd-$Z$ nuclei $^{45}$Mn, $^{49}$Co, $^{55}$Cu, $^{59}$Ga are proton-unbound.
$^{50}$Co is obtained to be weakly bound, although our large uncertainty does not exclude the possibility
that it is proton unbound.
The one-proton separation energy in $^{60}$Ga is obtained to be zero with an uncertainty of 180~keV, however.
At the same time, $^{54}$Cu is expected in our model to be proton-unbound with $|S_p|<1$~MeV.



For even-$Z$ $^{49}$Ni, $^{55}$Zn, $^{60}$Ge, $^{64}$Se, $^{68}$Kr, $^{73}$Sr  
we find $S_p >0$, while $S_{2p}<0$, however, these $|S_{2p}|$ values stay rather small.
Indeed, all of these nuclei have been observed~\cite{Giov2001,Stolz2005,Blank2016,Sinclair2019}. 
Some of them have tentatively been proposed as potential candidates for two-proton radioactivity according
to experimental indications, while $\beta $-decay might still be the preferred decay mode for a few of them 
(e.g., for $^{49}$Ni~\cite{Giov2001}, $^{68}$Kr~\cite{Giov2019} and $^{73}$Sr~\cite{Sinclair2019}).
Their neighbors with one neutron less,  $^{48}$Ni, $^{54}$Zn, $^{59}$Ge, $^{63}$Se, $^{67}$Kr, $^{72}$Sr, 
as well as $^{45}$Fe, are calculated to have a more pronounced and negative $S_{2p}$ and thus may
be considered as possible candidates for $2p$ emission (confirmed for $^{45}$Fe~\cite{GiovPRL2002,Pfutzner2002},
$^{48}$Ni~\cite{Dossat2005,Pomorski2011}, $^{54}$Zn~\cite{BlankPRL2005} and $^{67}$Kr~\cite{GoigouxPRL2016}). At the same time,
$^{59}$Ge and $^{63}$Se have been observed to disintegrate via $\beta $-decay~\cite{GoigouxPRL2016} which may not be surprising,
since the predicted $|S_{2p}|$ value is smaller than the one found for $^{67}$Kr
(see also a recent study from $^{78}$Kr fragmentation reported in Ref.~\cite{Blank2016}).

For heavier odd-$Z$ nuclei, we report on the possibility of negative $S_p$ values for 
$^{64,65}$As, $^{68,69}$Br, $^{71,72,73}$Rb, which is in accord with experimental studies
of those nuclei~\cite{TuPRL2011,RogersPRL2011,DelSanto2014,Suzuki2017,Sinclair2019}.
The less negative $S_p$ values for $^{64,65}$As are in agreement with the fact that 
those nuclei have been observed experimentally~\cite{BlankPRL1995,TuPRL2011}, 
while $^{69}$Br and $^{73}$Rb, having more negative $S_p$ values, were not observed~\cite{RogersPRL2011,DelSanto2014,Suzuki2017} and, 
therefore, may be proton-unbound with a very short half-life.
Among these nuclei are the $rp$-process waiting-point neighbors, $^{65}$As, $^{69}$Br and $^{73}$Rb,
which have thus important implication for astrophysics simulations.

We notice another interesting feature for a few series of odd-$Z$ isotopes, namely, 
we indeed observe the existence of the so-called ``sandbanks''~\cite{Suzuki2017}: 
$S_p$ for an even $A$ nucleus can be similar or even less negative
than for an odd-mass $(A+1)$ nucleus.
It is well illustrated by numerous examples from our chart, e.g., $^{68}$Br -- $^{69}$Br,
$^{66}$Br -- $^{67}$Br, $^{72}$Rb -- $^{73}$Rb,  $^{70}$Rb -- $^{71}$Rb, and heavier $^{80}$Nb -- $^{81}$Nb, etc.

For $^{76,77}$Y, our $S_p$ values stay small, but negative, within relatively small error bars (see Fig.~\ref{fig:chart2}), 
which is not completely excluded by the results on beta-decay half-lives of these isotopes 
deduced from the experimental studies~\cite{Faestermann2002,Sinclair2019}.
At the same time, we remark a very good agreement with the experimental indications for 
$^{81}$Nb, $^{85}$Tc, $^{89}$Rh, $^{93}$Ag and $^{97}$In to be proton unbound (see Ref.~\cite{Faestermann2002,CelikovicPRL2016}).
In particular, we find that our theoretical one-proton separation energies for $^{89}$Rh, $^{93}$Ag and $^{97}$In are close to experimentally 
determined values~\cite{CelikovicPRL2016}.
Similarly, we confirm a general tendency of a decrease of $|S_p|$ and, therefore, of an increase
of stability, when approaching $Z=50$ (see a slight change in the average value of $S_p$ when going 
from $^{85}$Tc to $^{89}$Rh, $^{93}$Ag and further to $^{97}$In). 
These are the nuclei which result from a proton-capture on even-even $N=Z$ nuclei (possible waiting points).

For even-$Z$ Sr and Zr, according to our calculation, $^{73}$Sr and $^{77}$Zr are slightly unbound, 
with the $S_{2p}$ value being close to zero. 
At the same time, we find $^{82}$Mo slightly bound with respect to two-proton emission, with a large uncertainty, however.

For the heavy even-even nuclei $^{86}$Ru, $^{90}$Pd, $^{94}$Cd, $^{98}$Sn, we get $S_p>0$ and $S_{2p}<0$, with $S_{2p}$ being very small.
Due to this reason, we would be cautious to provide any definite answer about the exact position of the two-proton drip-line here.
Their neighbors with one neutron less, $^{85}$Ru, $^{89}$Pd, $^{93}$Cd, $^{97}$Sn are predicted
to have $-1.5 \lesssim S_{2p}\lesssim -1$~(MeV), and thus are expected be two-proton unbound.
However, one might need to go to still more exotic isotopes, $^{84}$Ru, $^{88}$Pd, $^{92}$Cd or $^{96}$Sn,
in a search for plausible candidates for $2p$ emission.

We remark that the masses of heaviest $T_z=-1/2$ nuclei from $^{89}$Rh to $^{99}$Sn are based on the extrapolated 
masses of their $T_z=1/2$ mirrors~\cite{AME2016}.
Beyond Sn isotopes, we cannot approach the $N=Z$ line with our calculation, since the mass excesses of neutron-rich mirrors
become unknown.

Addition of the proper shell correction, capable to reproduce variations of $b$ coefficients in a better way, 
would in principle reduce the uncertainty on the proton and two-proton separation energies, as well as it could modify
their central values.\\


\section{Conclusions and summary}


In the present work, we have applied the macro-microscopic approach of P.~M\"oller and collaborators, the FRDM and FRLDM, 
to estimate the IMME $a$ and $b$ coefficients in nuclei in the vicinity of $N=Z$ until $A\sim 100$.
We found that the macroscopic part of the FRLDM, representing a refined version of the well-known liquid-drop model, 
is rather well suited to describe the general trends of the coefficients.
In particular, introduction of proton and neutron pairing energy parameters adjusted to experiment allowed us 
to reproduce the staggering effect of $b$ coefficients in very good agreement with the data.

The full FRDM and FRLDM approaches prove to be advantageous in providing the general trend of the $a$ coefficients towards heavier nuclei 
along the $N=Z$ line which would be interesting to confirm experimentally. However, the description of $b$ and $c$ coefficients is not satisfactory. 
This may hint at a possible opportunity to improve the parametrizations with a specific consideration of the isovector and isotensor 
parts of the total energy.

The ``best'' set of $b$ coefficients obtained from the modified macroscopic part of FRLDM is used to calculate 
the masses of proton-rich nuclei based on experimental masses of their neutron-rich mirrors and theoretical $b$ coefficients. 
The results obtained on one- and two-proton separation energies of nuclei in the vicinity of the proton drip-line are in robust
agreement with the available experimental data. The predictions made for heavier nuclei near $^{100}$Sn can serve to guide future experiments.

We hope that our study can motivate further exploration and developments of the FRDM and FRLDM, aiming at a better description
of the isospin degree of freedom, which could in future provide the community with more accurate predictions of masses of proton-rich nuclei. 
The properties of those nuclei up to $A\approx 100$ are very important for simulations of the astrophysical $rp$-process. \\

\acknowledgments{
We thank Bertram Blank for a careful reading of the manuscript and many valuable remarks.
The work was supported by CNRS/IN2P3, France, under Master Project ``Isospin''. 
O. Klochko thanks the CENBG for its hospitality during his three internship stays in 2018 and 2019.}

\bibliography{imme}

\end{document}